\documentclass[twocolumn,fleqn]{article}
\usepackage{geometry}
\usepackage{graphicx}
\usepackage{amsmath,amssymb,amsfonts}
\usepackage{xcolor}

\usepackage{caption}
\usepackage{subfig}
\usepackage{adjustbox}
\usepackage{multirow}

\newtheorem{theorem}{Theorem}
\newtheorem{lemma}[theorem]{Lemma}
\newtheorem{pf}{Proof}
\newtheorem{definition}{Definition}
\newtheorem{property}{Property}

\usepackage[ruled, lined, noend]{algorithm2e}
\SetKwRepeat{Do}{do}{while}
\SetKw{Return}{return}
\SetKw{Continue}{continue}
\SetKw{Break}{break}
\SetKwProg{Fn}{Function}{}{}
\SetKwFunction{func}{connect}
\SetKwFunction{Connected}{connected}
\SetKwFunction{FindMax}{findMax}
\SetKwFunction{Cut}{cut}
\SetKwFunction{Link}{link}

\title{Dynamic data summarization for hierarchical spatial clustering}
\author{
    Kayumov Abduaziz \thanks{
        Department of Computer and Information Security, and Convergence Engineering for Intelligent Drones,
        Sejong University,
        abduaziz@sju.ac.kr
    }
~~~~
    Min Sik Kim \thanks{
        Petabi Inc,
        msk@petabi.com
    }
~~~~
    Ji Sun Shin \thanks{
        Department of Computer and Information Security,
        Sejong University,
        jsshin@sejong.ac.kr
    }
}
\date{November 2024}

\begin{document}
\maketitle

\begin{abstract}
Hierarchical Density-Based Spatial Clustering of Applications with Noise (HDBSCAN) finds meaningful patterns in spatial data by considering density and spatial proximity.
As the clustering algorithm is inherently designed for static applications, so have recent studies focused on accelerating the algorithm for static applications using approximate or parallel methods.
However, much less attention has been given to dynamic environments, where even a single point insertion or deletion can require recomputing the clustering hierarchy from scratch due to the need of maintaining the minimum spanning tree (MST) over a complete graph.
This paper addresses the challenge of enhancing the clustering algorithm for dynamic data.
We present an exact algorithm that maintains density information and updates the clustering hierarchy of HDBSCAN during point insertions and deletions.
Considering the hardness of adapting the exact algorithm to dynamic data involving modern workloads, we propose an online-offline framework.
The online component efficiently summarizes dynamic data using a tree structure, called \textit{Bubble-tree}, while the offline step performs the static clustering.
Experimental results demonstrate that the data summarization adapts well to fully dynamic environments, providing compression quality on par with existing techniques while significantly improving runtime performance of the clustering algorithm in dynamic data workloads.
\end{abstract}

\section{Introduction}
\label{intro}
Hierarchical spatial clustering algorithms \cite{optics, hdbscan} are essential for analyzing spatial data as they allow for the identification of nested clusters and relationships at multiple density levels .
These algorithms build a hierarchy of clusters by successively merging smaller clusters or splitting larger ones \cite{hdbscan}, which is particularly useful when dealing with complex spatial datasets.
This multi-density approach is invaluable in unsupervised or semi-supervised learning, where understanding the spatial arrangement of points is crucial for geo-spatial clustering \cite{geo_spatial_clustering}, resource allocation \cite{resource_allocation_hdbscan}, real-time change detection of point cloud models \cite{lidar_hdbscan}, and log-based anomaly detection \cite{failure_detection_optics}.
Many of these applications also have their dynamic variants, requiring the continuous maintenance of hierarchical spatial clusters in response to often subtle changes in dynamic data.

While hierarchical spatial clustering algorithms have been studied well as static algorithms, recently, adapting them to such streaming or fully dynamic applications have attracted a great deal of research.
As their non-hierarchical algorithm has to resort to approximation techniques \cite{dynamic_dbscan}, their modification for hierarchical version is not a trivial problem.
Such an exact version of dynamic hierarchical spatial clustering algorithm can still be developed using dynamic index structures \cite{rtree} utilizing k-nearest neighbor queries \cite{knn} for density computation and reverse k-nearest neighbor queries \cite{rknn} for keeping track of affected clustering regions in the clustering hierarchy.
Although this is practically feasible, one has to deal with the inefficiencies introduced by the curse of dimensionality \cite{curse_of_dimensionality} and updating the clustering hierarchy (e.g. dendrogram or reachability plots) in the transformed distance space of the hierarchical clustering algorithm \cite{ihastream}, raising the theoretical guarantees to be close to re-computing the clustering hierarchy using the static algorithms.
Furthermore, running hierarchical spatial clustering algorithms on very large datasets for a long time is very counter practical, since such clustering algorithms are mainly used as a quick preprocessing step in data mining applications \cite{flame_using_hdbscan}.
Considering these reasons, data summarization techniques are employed to mitigate the complexity of fully dynamic hierarchical spatial clustering algorithms, providing more flexibility in selecting the desired level of compression while utilizing the existing static algorithm \cite{data_bubbles}.

Data summarization techniques have been used in hierarchical clustering to enhance scalability with minimal loss in clustering quality.
A renowned data summarization technique, called clustering features (CF) \cite{birch}, has been well-studied in the context of stream clustering algorithms, most prominent algorithms being density-based clusters \cite{denstream} as well as for anytime capabilities \cite{clustree}, sliding windows \cite{streamsw}, and online-offline framework \cite{online_offline_framework}.
These studies focus on maintaining micro-clusters derived from clustering features that are configured by a user-defined threshold radius parameter, such that a micro-cluster absorbs data points if the distance between the center of the micro-cluster and data points is within the threshold value.
One such dynamic case involves a sliding-window model that maintains micro-clusters for hierarchical spatial clustering \cite{hdbscan_sliding_window}.
For hierarchical clustering, clustering features are not directly applicable due to the imbalance in the number of data points represented by them, and less information on approximating the distances between original points, which are the main reasons that the clustering quality has been known to deteriorate rapidly when used as data summarization for hierarchical clustering \cite{data_bubbles}.
To remedy these issues, a post-processing step, called data bubbles \cite{data_bubbles}, has been introduced to better reflect the relationships between original points.
The effectiveness of data bubbles has been shown in the well-known hierarchical spatial algorithms such as Ordering Points To Identify the Clustering Structure (OPTICS) \cite{optics} and HDBSCAN (Hierarchical Density-Based Spatial Clustering of Applications with Noise) \cite{hdbscan} where data bubbles have been applied to approximate hierarchical clusters on very large datasets.
Combining data bubbles and MapReduce framework \cite{mapreduce}, the static algorithm of HDBSCAN has been extended to support parallel and distributed computation, bringing down months of computation to several hours with little loss in clustering quality \cite{hdbscan_map_reduce}.
For the realization of fully dynamic hierarchical spatial clustering, we would like an efficient and scalable data summarization that adapts well to the nature of dynamic data. 
Although, the maintenance of fixed number of data bubbles has been proposed \cite{maintain_fixed_data_bubbles} which has been further improved upon for dynamic number of data bubbles \cite{maintain_dynamic_data_bubbles},
the scalability challenges that emerge when the dataset size grows unpredictably in fully dynamic settings have not been considered further.

In this paper, we are interested in enhancing hierarchical spatial clustering for fully dynamic data applications.
Firstly, we develop an exact algorithm that maintains the clustering hierarchy under point insertions and deletions.
Secondly, we empirically analyze that exactly solving the problem is not practical such that a rather small change in the data may require the same computational work as the static algorithm.
Thirdly, to mitigate the scalability challenges and adapt for rapidly scaling fully dynamic data, we propose an online-offline framework where the online component maintains a compressed form of the dynamic data using a tree structure called Bubble-tree for a desired compression factor and the offline component performs a post-processing step using data bubbles for computing the clustering hierarchy.
Finally, an extensive experimental analysis of the proposed algorithms was conducted to assess the scalability and quality of the clustering hierarchy in fully dynamic settings.

The subsequent sections of this paper are organized as follows.
Section 2 provides a formal review of hierarchical spatial clustering, emphasizing recent relevant techniques of data summarization for dynamic environment.
In Section 3, we define the problem and propose an exact dynamic algorithm for hierarchical spatial clustering followed by its feasibility analysis.
In Section 4, we introduce Bubble-tree for dynamic data summarization and its integration with the clustering algorithm.
Section 5 validates the results of our study in comparison to the static and exact dynamic version of the clustering algorithm.
Section 6 gives an overview of recent related work.
Finally, we discuss future research directions and conclude the paper.

\section{Background}
\label{sec:preliminaries}
We describe the basic primitives of HDBSCAN and related concepts for a fully dynamic settings subject to point insertions and deletions.
Table \ref{table:notations} provides a summary of notations used throughout the rest of this paper.

\begin{table}[t]
\caption{Notations}
\begin{adjustbox}{width=1\linewidth}
\renewcommand{\arraystretch}{1.6}
    \centering
    \begin{tabular}{|c|l|}
        \hline
        {\large Notation} & {\large Description} \\
        \hline
        \multirow{2}{*}{\large $minPts$}            & {\large HDBSCAN's density parameter, a minimum number of} \vspace{-0.5ex}\\
                                                    & {\large points required to form a dense region.} \\
        \hline
        {\large $d(p, q)$}                          & {\large Euclidean distance between points $p$ and $q$.} \\
        \hline
        \multirow{2}{*}{\large $cd(p)$}             & {\large The core distance value of point $p$, represents} \vspace{-0.5ex}\\
                                                    & {\large Euclidean distance to its $minPts$-th nearest neighbor.} \\
        \hline
        \multirow{2}{*}{\large $d_m(p, q)$}         & {\large The weight of an edge $(p, q)$, represents the mutual} \vspace{-0.5ex}\\
                                                    & {\large reachability distance between points $p$ and $q$.} \\
        \hline
        \multirow{2}{*}{\large $N_{minPts}(p)$}     & {\large A neighborhood set of point $p$, contains pointers to} \vspace{-0.5ex}\\
                                                    & {\large the nearest $minPts$ points to $p$.} \\
        \hline
        \multirow{2}{*}{\large $R_{minPts}(p)$}     & {\large A reverse neighborhood set of point $p$, where $q \in R_{minPts}(p)$} \vspace{-0.5ex}\\
                                                    & {\large if $p \in N_{minPts}(q)$.} \\
        \hline
    \end{tabular}
\end{adjustbox}
\label{table:notations}
\end{table}

\begin{figure*}
\centering
        \subfloat[Neighboorhood]{%
            \includegraphics[width=0.25\linewidth]{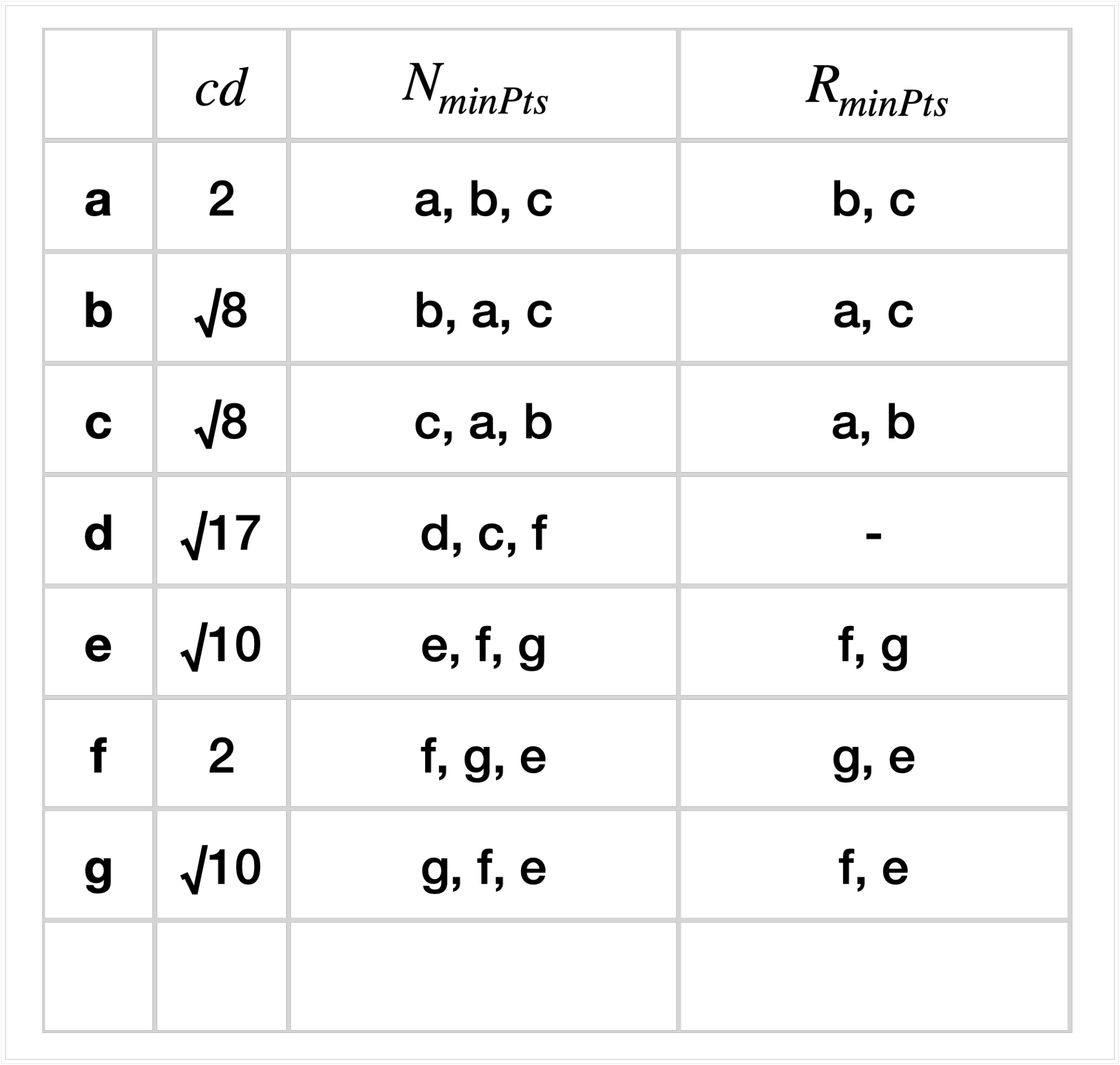}%
        }
        \subfloat[Distance matrix]{%
            \includegraphics[width=0.25\linewidth]{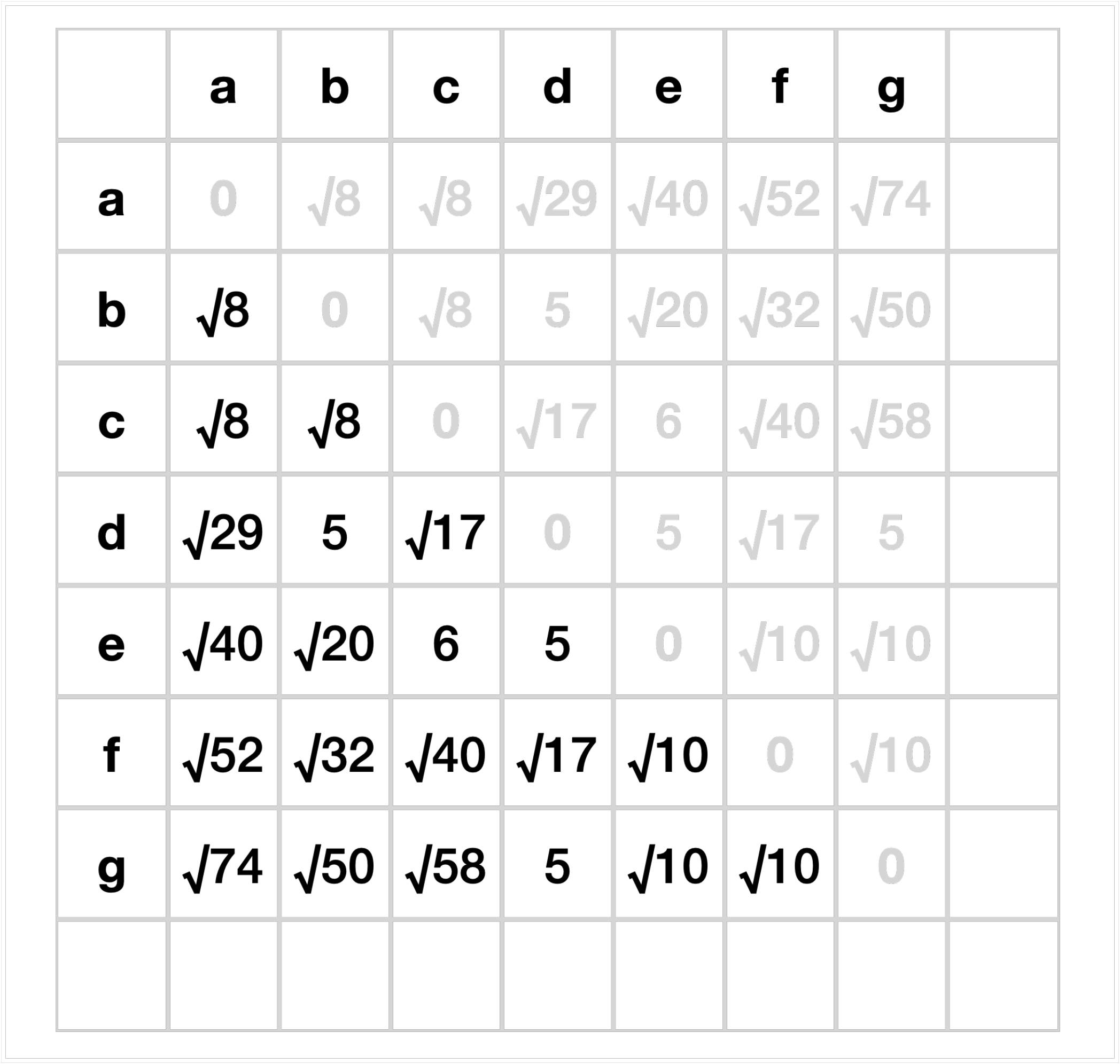}%
        }
        \subfloat[MST]{%
            \includegraphics[width=0.25\linewidth]{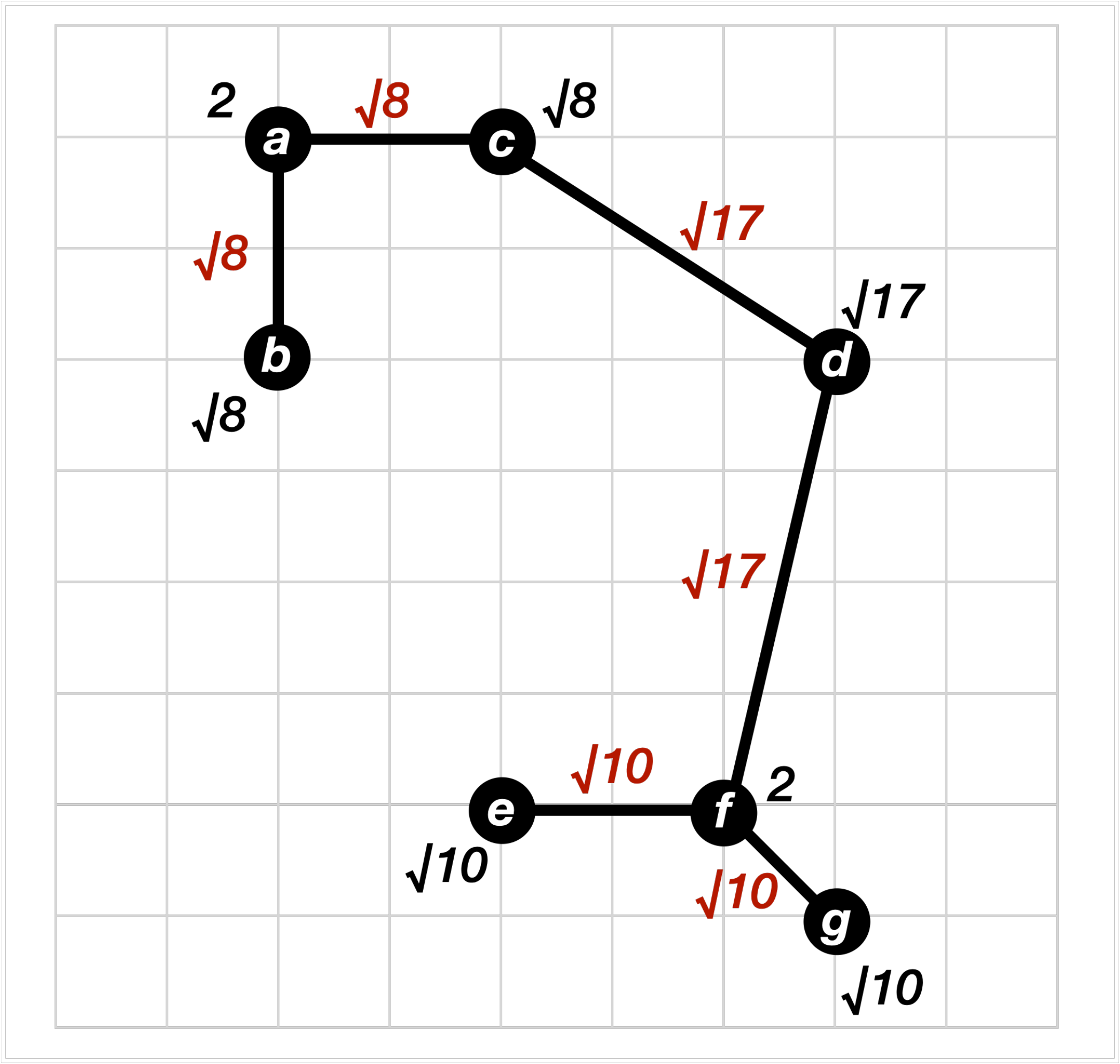}%
        }
        \subfloat[Dendrogram]{%
            \includegraphics[width=0.25\linewidth]{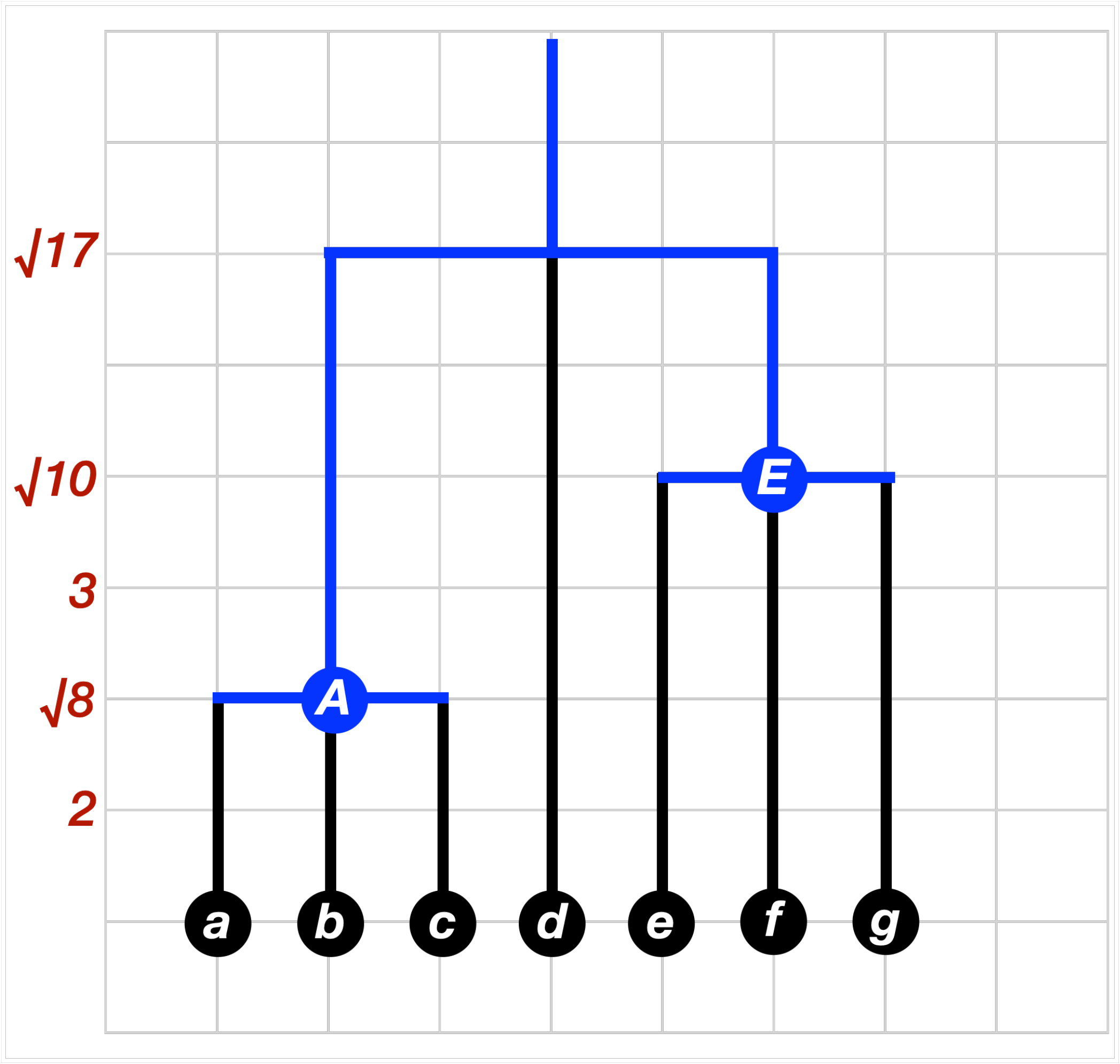}%
        }
        \caption{
        An illustration of HDBSCAN clustering results performed on 2D example data for $minPts = 3$.
        The neighborhood information (a) includes core distance values of data points computed using their nearest neighbors.
        The distance matrix (b) shows all pairwise mutual reachability distances, used as weights to compute the MST (c).
        The clustering hierarchy, called \textit{dendrogram} (d), is obtained from the MST by removing the edges in decreasing order of weights, resulting in two major clusters $A$ and $E$.
        }
        \label{fig:mst_example_before}
\end{figure*}

\subsection{Hierarchical spatial clustering}
HDBSCAN \cite{hdbscan} operates on the point set $P$ with a user-defined density parameter $minPts$ to extract a hierarchy of density-based clusters \cite{dbscan} at several density levels.
The most important concepts of core distance, mutual reachability distance between pairs of points and mutual reachability graph are defined as follows.

\begin{definition}[Core distance]
The \textit{core distance} of point $p$, denoted as $cd(p)$, is the distance to its $minPts$th nearest neighbor in $P$.
\end{definition}


\begin{definition}[Mutual reachability distance]
    The mutual reachability distance between two points $p$ and $q$, denoted as $d_m(p, q)$, is defined as the maximum of their core distances and the actual distance between them:
\begin{equation}
\label{eq:mreach}
    d_m(p,q) = \max \{ cd(p), cd(q), d(p, q) \}
\end{equation}
\end{definition}

\begin{definition}[Mutual reachability graph]
    The mutual reachability graph, denoted by $G$, is a complete undirected graph for the point set $P$ where the nodes represent the points in $P$ and the weights of all pairwise edges correspond to the mutual reachability distances between pairs of points.
\end{definition}

The clustering algorithm does not depend on a fixed radius threshold, $\epsilon$, needed for finding density-based clusters at density level $\epsilon$ \cite{dbscan}.
Instead, it explores all density-based clusters represented as a hierarchy of clusters in the transformed space of the mutual reachability distances through the following steps:
\begin{enumerate}
    \item {\it Tree construction}. Spatial tree structures, such as Ball-tree or kd-trees, are used to index the point set $P$ for efficient querying.
    \item {\it Core distances}. The core distances for all points in $P$ are computed by querying the spatial index for $minPts$ neighbors.
    \item {\it Minimum spanning tree}. A minimum spanning tree $T$ is computed from the mutual reachability graph $G$.
    \item {\it Clustering}. A clustering hierarchy is built by iteratively removing edges from the MST in decreasing order of weights, identifying major clusters using the connected components.
\end{enumerate}

An MST of HDBSCAN represents the clustering hierarchy of the data points at different proximity levels, thus obtaining all non-hierarchical density-based clustering results.
A popular MST algorithms can be employed such as Prim's MST on the mutual reachability graph.
Although the mutual reachability graph requires $O(n^2)$ space, the clustering algorithm does not necessarily need to store the graph as the mutual reachability distances between points can be computed on demand.
Hence, Dual-tree Boruvka-based algorithms \cite{dualtreeboruvka} enhance the computation of MST by efficiently employing space-partitioning trees such as Ball-tree and kd-trees \cite{accelerated_hdbscan}.
Figure \ref{fig:mst_example_before} illustrates HDBSCAN's clustering process for $minPts = 3$: (a) $minPts$-nearest neighbors $N_{minPts}$ of data points are used to compute core distance values, (b) the distance matrix represents pairwise mutual reachability distances, (c) the MST is computed over the distance matrix, and (d) the dendrogram shows hierarchical relationships to identify clusters at several density levels.

\subsection{Data summarization}
We generalize data summarization and its derivatives using the well known data summarization technique, clustering features from Balanced Iterative Reducing and Clustering using Hierarchies (BIRCH) \cite{birch}.
\begin{definition}[Clustering features, $CF$]
A clustering feature $CF$ represents a point set $P$ by a tuple $CF = \{ LS, SS, n \}$ where:
\begin{itemize}
    \item $LS = \sum p \in P$ (linear sum),
    \item $SS = \sum p^2 \in P$ (squared sum),
    \item $n = |P|$ (weight).
\end{itemize}
\end{definition}

The most important property of clustering features is the additivity theorem \cite{birch} where two clustering features $CF_i$ and $CF_j$ can be merged by adding their statistics to create another clustering feature:
\begin{equation}
    CF_i + CF_j = \{ LS_i + LS_j, SS_i + SS_j, n_i + n_j \}
\end{equation}
This allows us to fine tune the number of clustering features by merging some of them so that they satisfy memory or scalability constraints.
BIRCH builds a tree hierarchy of CFs, called CF-Tree, which is an iterative approach where it sets a threshold value $radius$ such that a point $p$ is absorbed by a leaf clustering feature $CF$ if their distance is less $d(p, rep) \leq radius$ where $rep = LS / n$.
While BIRCH initially works with a fixed threshold $radius$, it adapts when too many leaf \textit{CF}s are created by splitting nodes, compressing the tree, or rebuilding it with a larger threshold $radius$ value.

For hierarchical spatial clustering algorithms, the clustering features are prone to result in poor quality clustering results due to less information on estimating the distances between them.
Therefore, as a post-processing step, the concept of data bubbles \cite{data_bubbles} has been introduced to approximate hierarchical clustering with high quality clustering results.
Data bubbles have been used to scale up OPTICS \cite{optics} and for distributed and parallel computation of HDBSCAN \cite{hdbscan_map_reduce}.

\begin{definition}[Data bubble, $B$]
A data bubble $B$ represents a point set $P$ via a tuple $\{rep, n, extent, nnDist\}$ where the attributes of this tuple can be derived from the clustering features $CF = \{ LS, SS, n \}$ as follows:
\begin{align}
\label{eq:data_buble_equations}
&rep = LS / n \\
&extent = \sqrt{ \frac{2 \times n \times SS - 2 \times LS^2}{n \times (n - 1)} } \\
&nnDist(k) = {\left(\frac{k}{n}\right)}^{\frac{1}{d}} \times extent
\end{align}
\end{definition}

By using the representative object $rep$ instead of the original points, data bubbles condense a large dataset into a more compact form, addressing scalability issues on large datasets and enhance clustering quality \cite{data_bubbles}.
When applying the static clustering algorithm on data bubbles, the notions of $minPts$ neighborhood, core distances and mutual reachability distances are slightly modified to reflect their representative points for better distance estimation of data bubbles \cite{data_bubbles}.
Specifically, the core distance of each data bubble $B$ approximates the core distance of its representative point, $rep$, is computed as follows:
\begin{equation}
cd(B) = d(B, C) + C.nnDist(k)
\end{equation}
such that the combined weight of the data bubbles that are closer to $B$ than $C$ when added to $k$ represent $minPts$ original data points \cite{hdbscan_syed2015parallelization}.
Similarly, the mutual reachability distance between two data bubbles $B$ and $C$ also takes the core distance information into account:
\begin{equation}
    d_m(B, C) = \max \{ cd(B), cd(C), d(B, C)\}
\end{equation}
Following these slight modifications, the clustering hierarchy is computed using the static algorithm of HDBSCAN.
To extract flat clustering results from the hierarchy, a cluster is weighted by adding the weights of data bubbles \cite{hdbscan_syed2015parallelization}. 

For fully dynamic data, data bubbles vary in their representation of points, necessitating the ranking of data bubbles to redistribute lower-quality ones to conserve memory and perform the splitting of over-qualified ones.
The quality of a data bubble, called {\it data summarization index}, is determined by the proportion of points it represents in relation to the total number of points in the dataset \cite{maintain_fixed_data_bubbles} where the data summarization index $\beta(B)$ of a data bubble $B$ is defined as:
\begin{equation}
    \beta(B) = \frac{n}{N}
\end{equation}
where $N$ is the size of the whole dataset.
For a given set of data bubbles, the mean $\mu_\beta$ and standard deviation $\sigma_\beta$ of their data summarization indexes are computed, then a quality of data bubble $B$ is determined to be one of the following:
\begin{itemize}
\label{eq:bubble_groups}
    \item {\it "good"} if $\beta(B) \in [\mu_\beta - k \times \sigma_\beta, \mu_\beta + k \times \sigma_\beta]$
    \item {\it "under-filled"} if $\beta(B) < \mu_\beta - k \times \sigma_\beta$
    \item {\it "over-filled"} if $\beta(B) > \mu_\beta + k \times \sigma_\beta$
\end{itemize}
where $k$ can be computed from Chebyshev's Inequality for a desired probability of "good" quality data bubbles.
To maintain data bubbles in fully dynamic data applications, the goal is to increase the number of "good" quality data bubbles by splitting "over-filled" data bubbles or redistributing the points represented by "under-filled" data bubbles \cite{maintain_fixed_data_bubbles}.

\subsection{ClusTree.}
ClusTree \cite{clustree} is an extension of R-tree family \cite{rtree} index structures for stream clustering based on the CF tree.
Similar to dynamic index structures, the ClusTree is a balanced tree structure where instead of rectangles, it stores the clustering features on each tree node.
The insertion algorithm follows the standard dynamic index algorithm in that a new point traverses down in the tree hierarchy, updating the clustering features on each level \cite{birch}.
When a new point reaches a leaf node, the leaf node absorbs the new point if the distance between the leaf and the point is within the current threshold, otherwise a new leaf is created.

Common to stream clustering algorithms \cite{denstream}, ClusTree adopts the damped window model to forget "old" points with decaying function: $\text{CF}(t+\Delta t) = \omega(\Delta t) \cdot \text{CF}(t)$, 
where $\omega(\Delta t) = \beta^{-\lambda \Delta t}$ defines the decay function that gives higher priority to newer objects in the streaming environment.
This allows an efficient deletion of old data for streaming data.
To adapt to varying stream speeds, ClusTree also modifies the insertion algorithm to be stopped anytime, where the unfinished insertions are kept in the buffer of internal tree nodes.
The hitchhiking concept allows for the continuation of such unfinished insertion operations where the newly-inserted points carry the buffered objects along their insertion path.

\begin{figure*}
\centering
        \subfloat[Neighborhood]{%
            \includegraphics[width=0.25\linewidth]{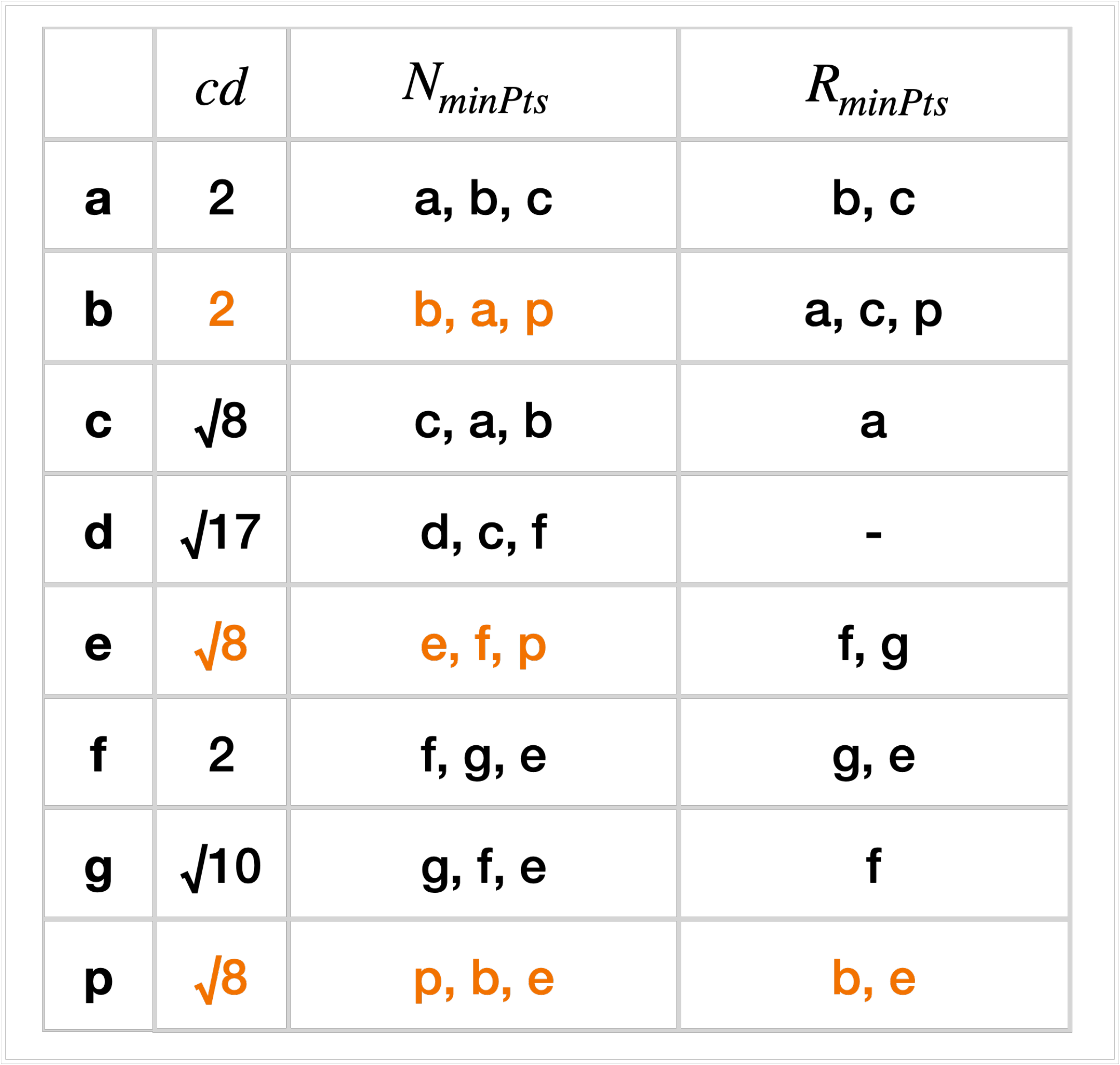}%
        }
        \subfloat[Distance matrix]{%
            \includegraphics[width=0.25\linewidth]{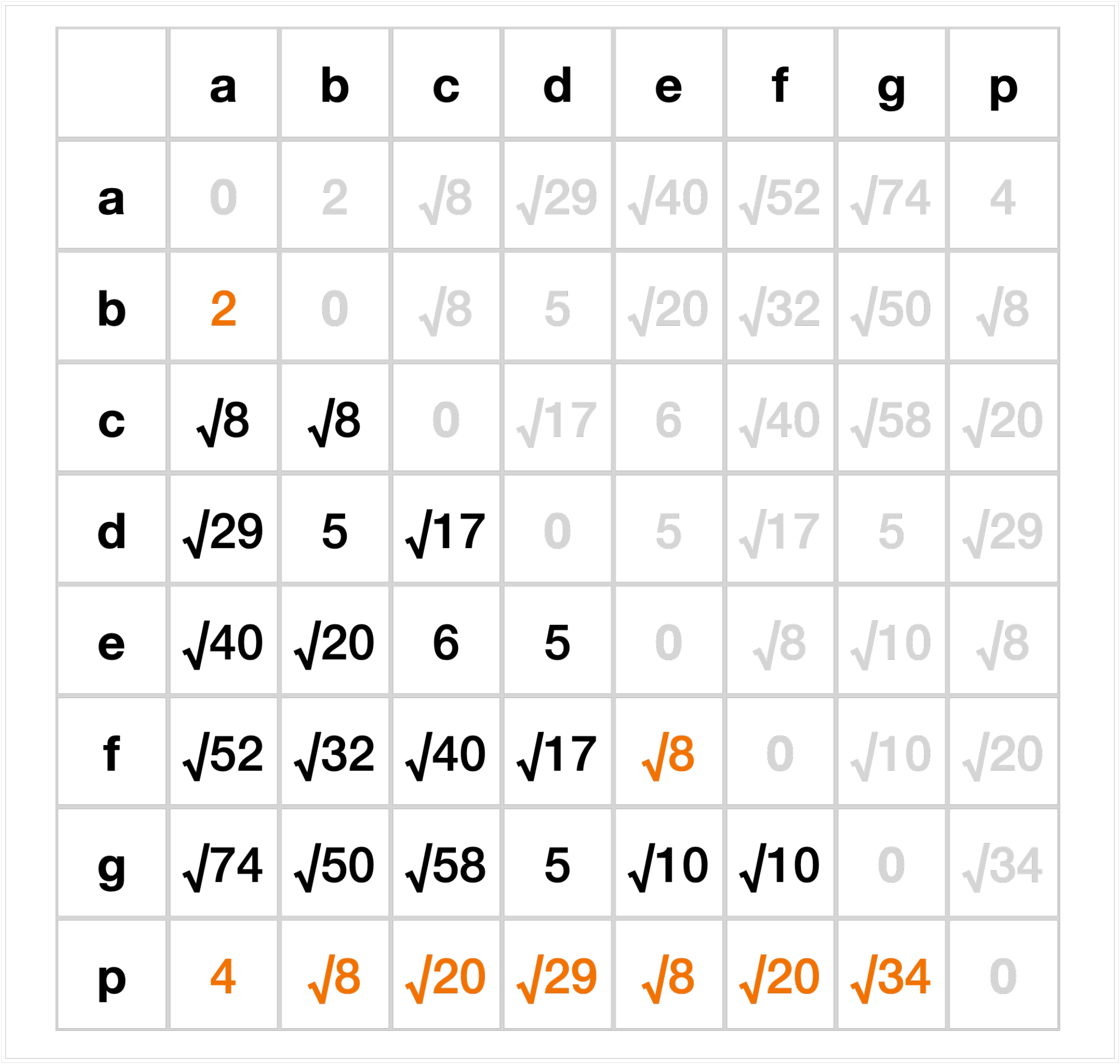}%
        }
        \subfloat[MST]{%
            \includegraphics[width=0.25\linewidth]{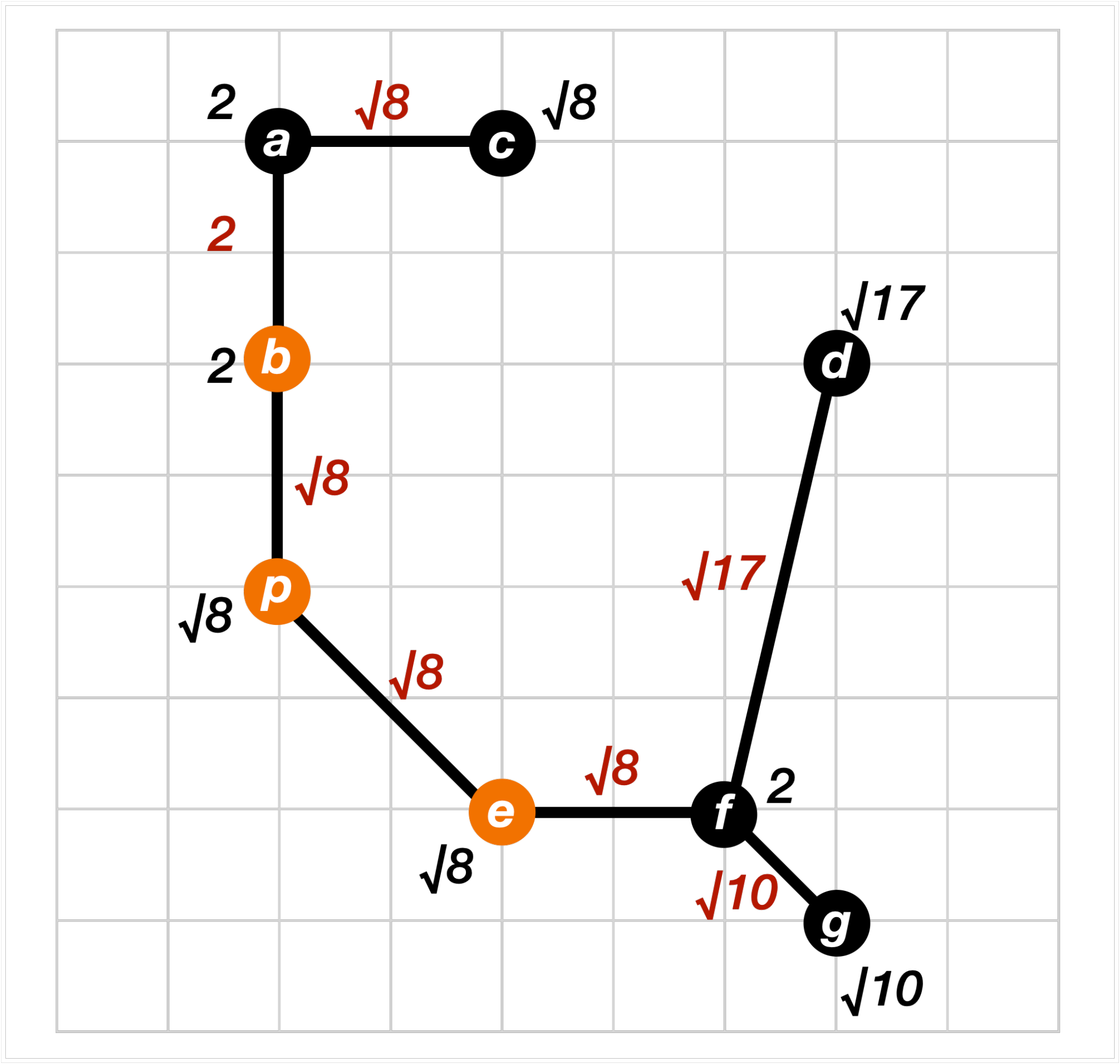}%
        }
        \subfloat[Dendrogram]{%
            \includegraphics[width=0.25\linewidth]{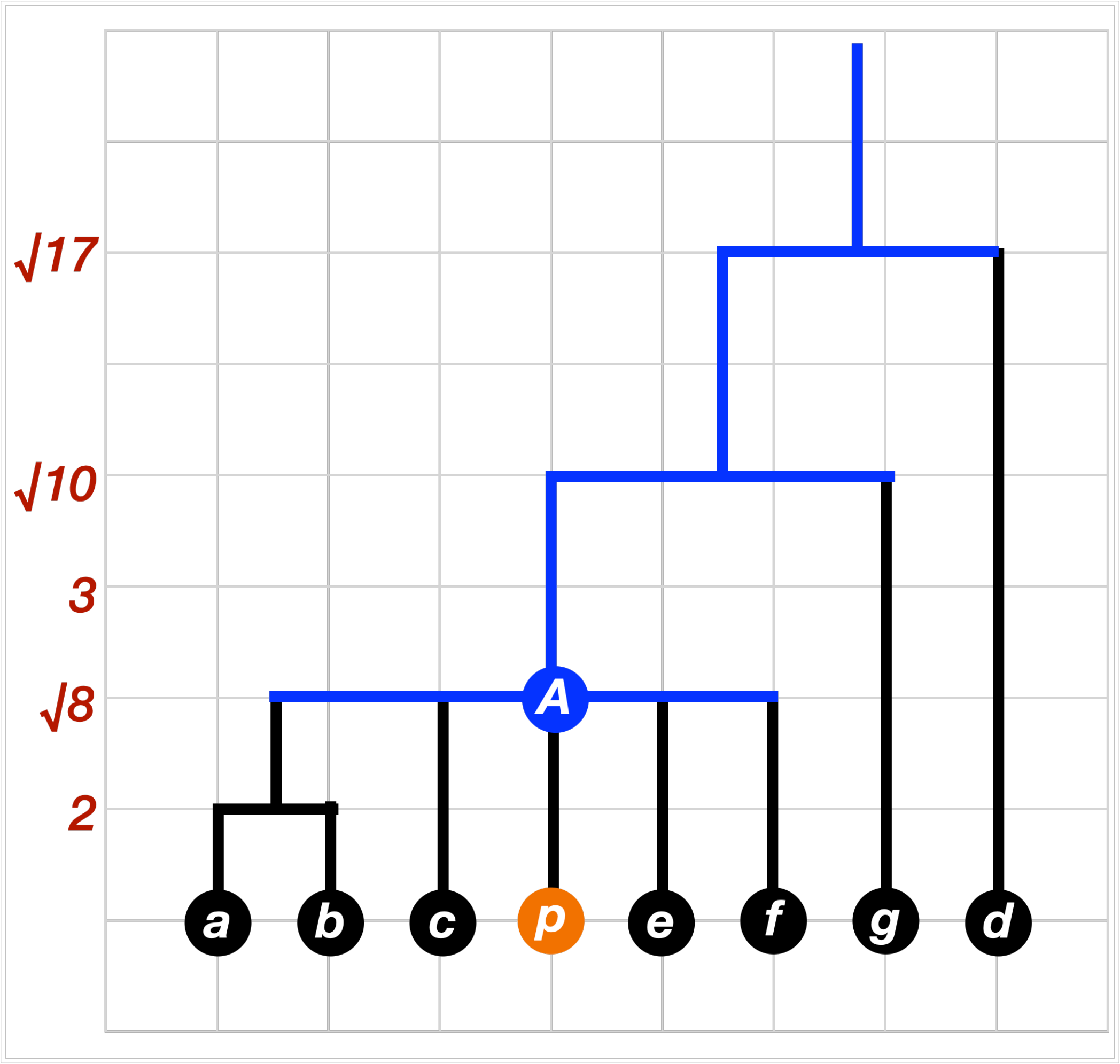}%
        }
        \caption{
        An illustration of HDBSCAN clustering results for $minPts = 3$ performed for the same 2D example data shown in Figure \ref{fig:mst_example_before} updated with the insertion of point $p$.
        The neighborhood information (a) highlights updated core distance values of points $b$ and $e$.
        The distance matrix (b) shows updated mutual reachability distances which are reflected in the MST (c).
        The clustering hierarchy shows an emergence of a single cluster (d).
        }
        \label{fig:mst_example_after}
\end{figure*}

\section{Dynamic hierarchical spatial clustering}
\label{sec:exact_algorithm}
\subsection{Problem definition}
In the context of hierarchical spatial clustering, consider a mutual reachability graph $G = (V, E)$ of $n$ data points that is transformed into $G' = (V', E')$ by a single insertion or deletion of point $p$.
Given the MST $T$ of $G$, the problem we aim to focus is the efficient maintenance of the MST $T'$ of $G'$.
The state-of-the-art algorithms operate on the whole dataset in that they compute $T'$ from scratch in $O(n \log n)$ expected time \cite{accelerated_hdbscan}.
We are interested in a practical solution that adapts to the changes in $G'$ to compute $T'$.

Due to the nature of mutual reachability graphs, we observe that $G'$ represents a slightly modified snapshot of $G$. 
Specifically, the insertion of $p$ decreases the core distances of a number of existing points in $G$, while the deletion of $p$ removes one of the nearest neighbors of some of the existing points in $G$.
Such points with modified core distances are referred as the {\it reverse k nearest neighbors} \cite{rknn} (RkNN) of $p$. 
Changes in the core distance values of RkNNs affect their corresponding edge weights (see Eq.\ref{eq:mreach}).
For an insertion, the RkNNs of $p$ have decreased core distances and may therefore offer lighter edges that can become part of $T'$.
On the other hand, RkNNs of $p$ may possess heavier edges in $T$ due to the increase in their core distances as $p$ is being deleted from $G$.
To summarize, we should solve the following sub-problems: performing efficient kNN and RkNN queries on the dynamic data for the dynamic maintenance of core distance information, and updating the MST to reflect the modified core distance information under point insertions and deletions.
Figure \ref{fig:mst_example_after} shows an updated HDBSCAN's clustering results for the same toy data given in Figure \ref{fig:mst_example_before}: (a) the neighborhood table highlights updated $minPts$-nearest neighbors $N_{minPts}$ of points $b$ and $e$ which are reverse $minPts$-nearest neighbors of the new point $p$,
(b) the distance matrix includes updated mutual reachability distances, (c) the MST is computed with respect to the updated weights, which translates to a significant change in the clustering hierarchy (d).

\subsection{Exact solution}
We present point insertion and deletion algorithms that maintains the core distance information of fully dynamic data and  the clustering hierarchy of HDBSCAN.
We assume an online phase where the dynamic maintenance of the MST of HDBSCAN is performed under point insertions and deletions, followed by the offline phase which involves extracting the clustering hierarchy from the MST at a user request.
We leverage a dynamic spatial index to keep track of modified core distances dynamically based on a reverse nearest-neighbor search \cite{rknn}.
The insertion algorithm uses dynamic trees \cite{link_cut_tree} to link and cut the edges modified by point insertions.
The deletion algorithm runs Boruvka's MST algorithm on the dynamic spatial index based on the dual-tree method \cite{dualtreeboruvka}.

Firstly, we shall define the difference between $G$ and $G'$ in terms of edge insertions and deletions as well as edge weight modifications:
\begin{equation}
    \Delta E = E_{inserted} \cup E_{deleted} \cup E_{modified}
\end{equation}

Next, we know that both $E_{inserted}$ and $E_{deleted}$ only contain such edges having point $p$ (inserted or deleted) as one of their endpoints.
It is also straightforward to derive that $E_{modified}$ represents such edges that are incident to RkNNs of point $p$ that connects them with their neighbors as the change in their core distances only affects their existing neighborhood set:
\begin{equation}
    E_{modified} = \{ (r, r') \mid r \in R_{minPts}(p) \; \wedge \; r' \in N_{minPts}(r) \}
\end{equation}

For clarity, we use $T$ to refer to the current MST, which is transformed into $T'$ by the insertion or deletion of point $p$.
We define the relationships between $T$ and $T'$ according to the reduction and contraction rules \cite{mst_reduction_and_contraction} reformulated for node-dynamic \cite{dynamic_graph_models} mutual reachability graph.

\subsubsection{Insertion}
The insertion algorithm is based on the link-cut tree algorithm presented in earlier work \cite{cattaneo2010maintaining}, where the insertion of point $p$ creates $n$ new edges that connect $p$ to all of the current points.
In order to transform $T$ into $T'$, we would like to probe the newly created edges and the existing edges with decreased weights incident to the RkNNs of $p$ into $T$, as they represent the difference between two snapshots of their underlying mutual reachability graphs.
To compute $T'$, we obtain the following formulation per reduction rule \cite{mst_reduction_and_contraction}:
\begin{equation}
\label{eq:mst_insertion}
    T' \subseteq T \cup E_{inserted} \cup E_{modified}
\end{equation}

Since the average number of RkNNs is closely related to the choice of $minPts$ \cite{knn_correlated_to_rknn}: $|E_{modified}| \approx minPts^2$ and $|E_{inserted}| = n$, the above formulation gives us a practically viable solution to update the clustering hierarchy of HDBSCAN under point insertions, compared to the static recomputation of the hierarchy with quadratic number of edges of the mutual reachability graph.

\subsubsection{Deletion}
The deletion of point $p$ increases the core distance values of its RkNNs. Accordingly, the initial step is to find the RkNNs of $p$ and re-compute their core distance values as explained in the insertion algorithm.
The increased core distance values of the RkNNs of $p$ may also increase the weights of their outgoing edges in the current MST.
To apply the contraction rule \cite{mst_reduction_and_contraction} we obtain the following forest $F$:
\begin{equation}
\label{eq:mst_deletion}
    F = T \setminus (E_{deleted} \cup E_{modified}) \subseteq T'
\end{equation}
The connected components of $F$ represent Boruvka components, therefore, we can run the dual-tree method \cite{dualtreeboruvka} to compute $T'$.
In the worst case, this approach may create $n$ connected components, essentially providing us with the recomputation of the MST from scratch.
However, the average number of RkNNs is closely related to the choice of $minPts$ \cite{knn_correlated_to_rknn}, hence we expect the algorithm to run on a much smaller number of disconnected components in practice.

\subsubsection{Implementation}
To realize an exact dynamic hierarchical spatial clustering, we assume an online maintenance of MST of HDBSCAN where the offline phase extracts the clustering hierarchy (dendrogram) at a user request.
We use the following data structures: a dynamic spatial index, SS-tree \cite{sstree} for maintaining core distance information and a link-cut tree \cite{link_cut_tree} for maintaining the MST.
Both insertion and deletion algorithms synchronizes the core distance information using kNN \cite{knn} and RkNN queries \cite{rknn} using a modified version of standard dynamic index operations \cite{rdnn}.
Appendix \ref{appendix:exact_dynamic} provides detailed pseudocode for the insertion and deletion algorithms.

\subsection{The hardness of exact algorithm}
Both insertion and deletion cases involves theoretically expensive maintenance of the MST of HDBSCAN as the underlying mutual reachability graph is a complete undirected graph.
The following lemmas give an overview of the hardness of maintaining the MST under point insertions and deletions.

\begin{lemma}
    Given an MST $T$ of a mutual reachability graph $G$, the insertion of point $p$ into $G$ requires $\Omega (n \log n)$ time to compute the updated MST $T'$ exactly.
\end{lemma}

\begin{pf}
    The insertion of point $p$ creates $n$ new edges, connecting $p$ to the existing points in $G$, thus, linearly expanding the graph under node-dynamic settings \cite{dynamic_graph_models}.
    Each of the newly-created edges can be a part of the updated MST $T'$, possibly replacing an existing edge in $T$ to satisfy the cut property of MST.
    To check if an edge $(p, q)$ is a part of $T'$, we need to confirm if points $p$ and $q$ are already connected in $T'$ to enforce the cycle property of MST, which takes $O(\log n)$ amortized time \cite{link_cut_tree}.
\end{pf}

\begin{lemma}
    Given an MST $T$ of a mutual reachability graph $G$, the deletion of point $p$ from $G$ requires $\Omega (n \log |F|)$ time to compute the updated MST $T'$ exactly where $F$ is the resulting forest from Equation \ref{eq:mst_deletion}.
\end{lemma}

\begin{pf}
    The deletion of point $p$ from $G$ also deletes all the edges incident to $p$.
    The most straightforward possibility is that $T$ might have contained only the edges incident to $p$, which in this case requires the recomputation of the updated MST $T'$ of $G'$.
    Next, the deletion of $p$ also requires the recomputation of core distance information of reverse nearest neighbors of $p$.
    Even worse, if $p$ happened to be the hub of the graph \cite{hubs_in_space}, we would have to recompute core distance values for extremely high number of reverse nearest neighbors of $p$, which requires $O(n^2)$ in the worst case.
\end{pf}

\begin{figure}[t!]
        \subfloat[]{%
            \includegraphics[width=.5\linewidth]{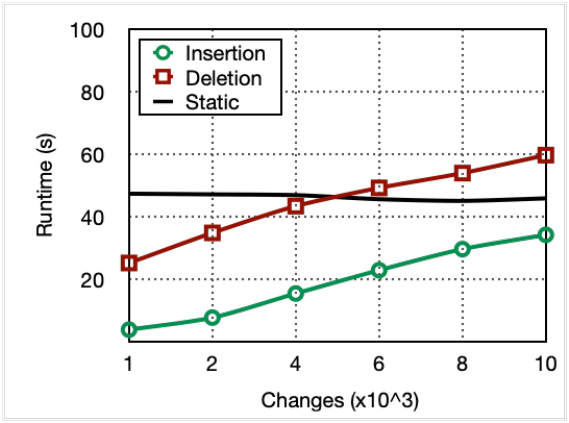}%
            \label{subfig:gauss1}%
        }
        \subfloat[]{%
            \includegraphics[width=.5\linewidth]{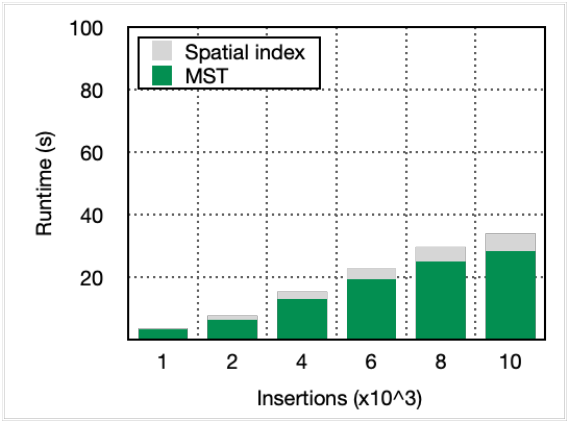}%
            \label{subfig:gauss2}%
        }
        \\
        \subfloat[]{%
            \includegraphics[width=.5\linewidth]{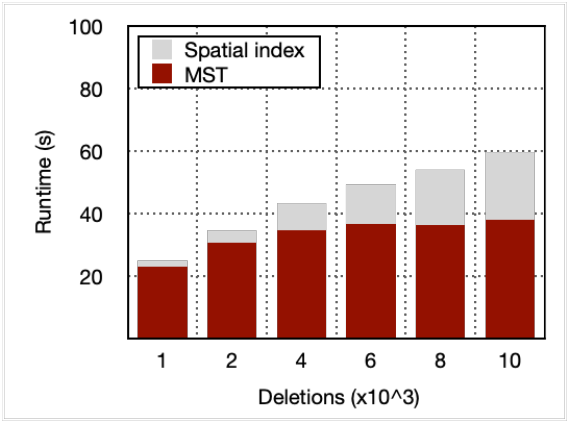}%
            \label{subfig:gauss3}%
        }
        \subfloat[]{%
            \includegraphics[width=.5\linewidth]{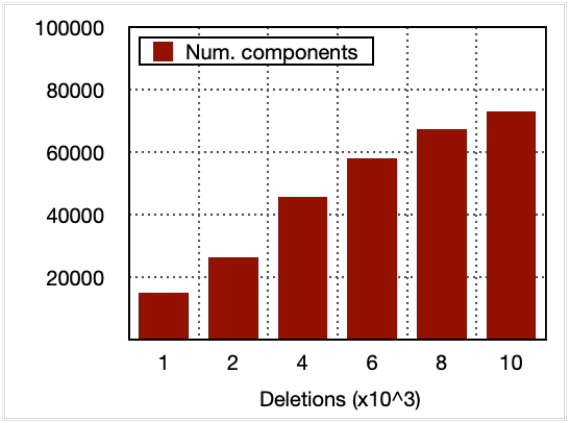}%
            \label{subfig:gauss4}%
        }
        \caption{
        Feasibility analysis of the exact dynamic algorithm on the Gaussian Mixtures dataset of 100K points for $minPts = 10$.
        }
        \label{fig:motivation}
\end{figure}

This analysis shows that even subtle minor changes in the data may cause the exact dynamic algorithm to underperform.
For further feasibility analysis of the exact dynamic algorithm, we used synthetic Gaussian Mixtures dataset \cite{mixsim} comprising 10 dimensional 100K points as given in Figure \ref{fig:motivation}.
Initially, we computed the exact algorithm for this dataset for $minPts = 10$, and then applied 1\%-10\% new insertions and deletions.
The runtime of each update was measured and compared to the static algorithm of HDBSCAN from the scikit learn library \cite{scikit}.
As shown in Figure \ref{subfig:gauss1}, the running time of the exact algorithm grows dramatically as more insertions and deletions were performed.
When more than 6\% deletions were applied, the performance of the exact algorithm was worse than the static algorithm.
To further analyze the slow performance of the exact algorithm, we measured the runtime decomposition as shown in Figures \ref{subfig:gauss2}-\ref{subfig:gauss3}.
As expected, the runtime of updating the MST was the dominant factor for both insertions and deletions.
This is due to the change in core distance information: as more updates are performed, large portions of the existing points have to recompute their core distances, so the MST of HDBSCAN should be updated to reflect the changes.
This effect is especially evident from the deletion case as shown in Figure \ref{subfig:gauss4}: as more deletions are performed, more Boruvka components have to be created to reflect the change in core distance information, thus proportionally growing its runtime.
The experiment demonstrates that the exact algorithm is not suitable for modern fully dynamic workloads: a small change in the data may dramatically increase its runtime, making the static algorithm a preferred option that runs on a bulk-loaded spatial tree \cite{accelerated_hdbscan} having performance advantages over dynamic spatial tree \cite{str}.
To mitigate the scalability issues, we consider data summarization techniques in the next sections.

\section{Proposed approach}
\label{sec:proposed_approach}
As the exact dynamic approach described earlier has scalability limitations in that rather small changes in the dynamic data may require the same amount of work of recomputing the clustering hierarchy of HDBSCAN due to the maintenance of its MST.
To mitigate this issue, we consider data approximation techniques where the online phase dynamically maintains a compressed form of the dynamic data, and the offline phase runs the static clustering algorithm on the compressed data.
For dynamic data compression, we design a dynamic tree structure called {\it Bubble-tree} with node entries containing clustering features for efficient update operations of fully dynamic data.
The core idea is that a clustering feature shares similarities with {\it minimum bounding sphere} employed in SS-Tree \cite{sstree}; therefore, we tailor well-established dynamic index algorithms to suit clustering features.
We also employ a quality measure from \cite{maintain_fixed_data_bubbles} to maintain high-quality clustering features w.r.t. the desired compression factor.
For the offline phase, we perform the static algorithm on the data bubbles \cite{data_bubbles} obtained from the clustering features of the leaf level of Bubble-tree.

\subsection{Bubble-tree}
\label{sec:bubble_tree}
Let $m$ and $M$ be the minimum and maximum fanout parameters, respectively, where $2 \times m \leq M + 1$, Bubble-tree is a fully dynamic balanced tree structure of clustering features that summarizes $N$ points with $L$ clustering features, satisfying the following properties:

\begin{property}
\label{property:root}
The root node has between two to $M$ children and stores a clustering feature representing the entire dataset.
\end{property}

\begin{property}
\label{property:internal_node}
An internal tree node stores pointers to between $m$ to $M$ tree entries (unless it is the root node) and represents them as a clustering feature.
\end{property}

\begin{property}
\label{property:leaf}
A leaf node's clustering feature represents actual points while an internal node stores a clustering feature representing the clustering features of its children.
\end{property}

\begin{property}
\label{property:compression}
The number of leaf nodes are maintained to be $L$ under point insertions or deletions.
\end{property}

While the fanout parameters $m$ and $M$ maintains Bubble-tree as balanced, $L$ leaf level nodes represent clustering features summarizing the fully dynamic data.
Here, $L$ can be configured to a desired compression factor, such as 10\% of $N$.
This hierarchical arrangement of clustering features enables efficient point insertions and deletions while maintaining desired number of clustering features of the dynamic data.
The insertion algorithm follows a standard depth-first strategy from dynamic index structures, where a newly-arrived point updates their clustering features of the nodes as it traverses down in the tree hierarchy until it reaches a leaf node which will become its parent.
The deletion algorithm deletes the point from its parent and its ancestors, if the leaf node representing the deleted point has less than $m$ children, the leaf is deleted and all of its remaining $m$ children are reinserted back into the tree.

After the insertion or deletion of point is performed on Bubble-tree, the maintain compression function is called to create or delete leaf nodes, as described in Algorithm \ref{alg:maintain_compression}.
The \texttt{MaintainCompression()} function operates on a given Bubble-tree $B$ with the compression factor $L$.
We aim to maintain the number of leaf nodes of Bubble-tree to be $L$.
If the current number of leaf nodes is larger than $L$, meaning that we are over-representing the data for the compression factor, the function proceeds by deleting the most under-filled leaf node from the tree and reinserting its children back to tree (Lines 2--4).
Another case is the current number of leaf nodes is smaller than $L$, meaning that we are under-representing the data, the function chooses the most overfilled leaf and splits it to create its sibling which will be reinserted back into the tree (Lines 6--8).
Here, both of the cases involve reinsertion of tree entries to enable dynamic reorganization of the tree as seen in the literature \cite{rstar}.
The split algorithm simply selects two farthest pairs among the tree node's children as seeds, reallocating the children to either seed to ensure each group contains at least $m$ children.
Bubble-tree additionally reorganizes current data summarization as the dynamic data evolves (Lines 10--11).

The maintenance of balanced-weighted clustering features in Bubble-tree w.r.t. the current number of points and $L$ is central to the purpose of this paper, and follows the well-studied data bubble compression quality metrics \cite{maintain_fixed_data_bubbles}, the difference being that Bubble-tree is a tree structure to allow rapid update operations on clustering features as the tree height is balanced under the fanout parameters of internal nodes.
Compared to ClusTree \cite{clustree}, Bubble-tree is not based on the decay function, as it targets fully dynamic data applications of data summarization.

\SetCommentSty{mycommfont}
\begin{algorithm}
\caption{Maintaining the compression factor of Bubble-tree}
\label{alg:maintain_compression}
    \KwIn{Bubble-tree $B$, $L$}
    \KwOut{Bubble-tree $B$}
        \nl \uIf{$B.num\_leaves > L$}{ \tcp{Fix under-representation}
                \nl Choose the most underfilled leaf $U$ \\
                \nl Remove $U$ from its ancestors in $B$ \\
                \nl Reinsert the points of $U$ into $B$
            }
        \nl \uElseIf{$B.num\_leaves < L$}{ \tcp{Fix over-representation}
                \nl Choose the most overfilled leaf $O$ \\
                \nl Split $O$ to create a sibling node $O'$ \\
                \nl Reinsert node $O'$ into $B$
            }
        \nl \uElse { \tcp{Dynamic reorganization steps} 
                \nl Choose the most overfilled leaf $O$ \\
                \nl Extract and reinsert $m$ farthest children of $O$ \\
            }
\end{algorithm}

\subsection{Clustering}
\label{sec:proposed_clustering}
At any given time, a hierarchical clustering algorithm can be directly applied to the clustering features of the leaf nodes of Bubble-tree, which are of size $L$, instead of $N$, leading to a significant reduction in size.
The process of integrating Bubble-tree and HDBSCAN for dynamic data involves the following steps:
\begin{itemize}
    \item \textit{Dynamic data summarization.}
    Perform point insertion and deletions on Bubble-tree.
    At a user request, extract $L$ clustering features from the leaf level of Bubble-tree and perform the following offline steps.
    \item \textit{Pre-processing.}
    Compute $L$ data bubbles from the clustering features obtained from the previous step.
    Assign the original data points to their closest data bubbles.
    \item \textit{Clustering.}
    Run the static algorithm of HDBSCAN on the data bubbles obtained from the previous step.
    Optionally, extract the most prominent flat clusters from the clustering hierarchy of HDBSCAN.
\end{itemize}

The dynamic data summarization step acts as an online component: point insertions and deletions are made on Bubble-tree for summarizing the dynamic data with a desired compression factor, whereas the pre-processing and clustering steps are offline components that are performed at a user request.

\section{Experiments}
\label{sec:experiments}
We conducted an experimental evaluation of \textit{Bubble-tree} for performing data summarization of dynamic data, focusing on both its performance and the quality of data compression it maintains.
The following data summarization algorithms were evaluated as the baseline algorithms:
\begin{itemize}
    \item {\bf Bubble-tree:} This is our Bubble-tree based approach where data summaries are obtained by the leaf level. We implemented Bubble-tree in Rust.
    \item {\bf ClusTree:} This is an incremental data summarization tree structure for stream clustering \cite{clustree}.
    We used ClusTree's implementation in R obtained from the streamMOA library \cite{streamMOA}.
    \item {\bf Incremental:} This refers to the incremental maintenance of data bubbles using data summarization index \cite{ maintain_dynamic_data_bubbles,maintain_fixed_data_bubbles}. We used our own implementation in Rust.
\end{itemize}

The baseline data summarization techniques were used as \textit{online} data summarization component, where the offline component involved running the static algorithm of HDBSCAN \cite{hdbscan}(implemented in Rust) on the summarized data as described in Section \ref{sec:proposed_clustering}.
We additionally measured the performance of the fastest implementation of HDBSCAN Python \cite{accelerated_hdbscan}, denoted as \textbf{Static}.
The exact dynamic algorithm proposed in Section \ref{sec:exact_algorithm} is referred as \textbf{Dynamic}.
For all of the baseline algorithms, the density parameter of HDBSCAN was configured to be $minPts = 100$ which results in 10-50 major clusters obtained from the clustering hierarchy.
We conduct experiments on a Macbook Pro (14-inch, 2021) with Apple M1 Pro and 16 Gb of RAM.
The experiments were performed using Rust's optimized "release" mode to ensure fair comparison of the baseline algorithms.

We generated a synthetic dataset, \textbf{Gauss}, using Gaussian Mixtures package \cite{mixsim}, consisting 10 dimensional 5,000,000 points, configured with a 10\% desired maximum overlap between gaussian mixture clusters.
The \textbf{Pamap} dataset \cite{pamap} consists of 4 dimensional 3,850,505 sensor measurements obtained after dimensionality reduction of human activity recognition data.
The \textbf{Chem} dataset \cite{UCI_chem} consists of 16 dimensional 4,178,504 recordings from 16 chemical sensors exposed to two dynamic gas mixtures with varying concentrations.
The \textbf{Intrusion} dataset \cite{kdd_cup_1999_data_130} comprises 34 dimensional 4,898,430 network connection logs, commonly used for identifying network intrusions.
We additionally used a 2D toy dataset of 1,000 points, \textbf{Seeds}, for visualization purposes.
All of the datasets above, even under their peak workload fit into the main memory of the experimental environment.

\begin{figure*}[t!]
\centering
        \subfloat[]{%
            \includegraphics[width=0.25\linewidth]{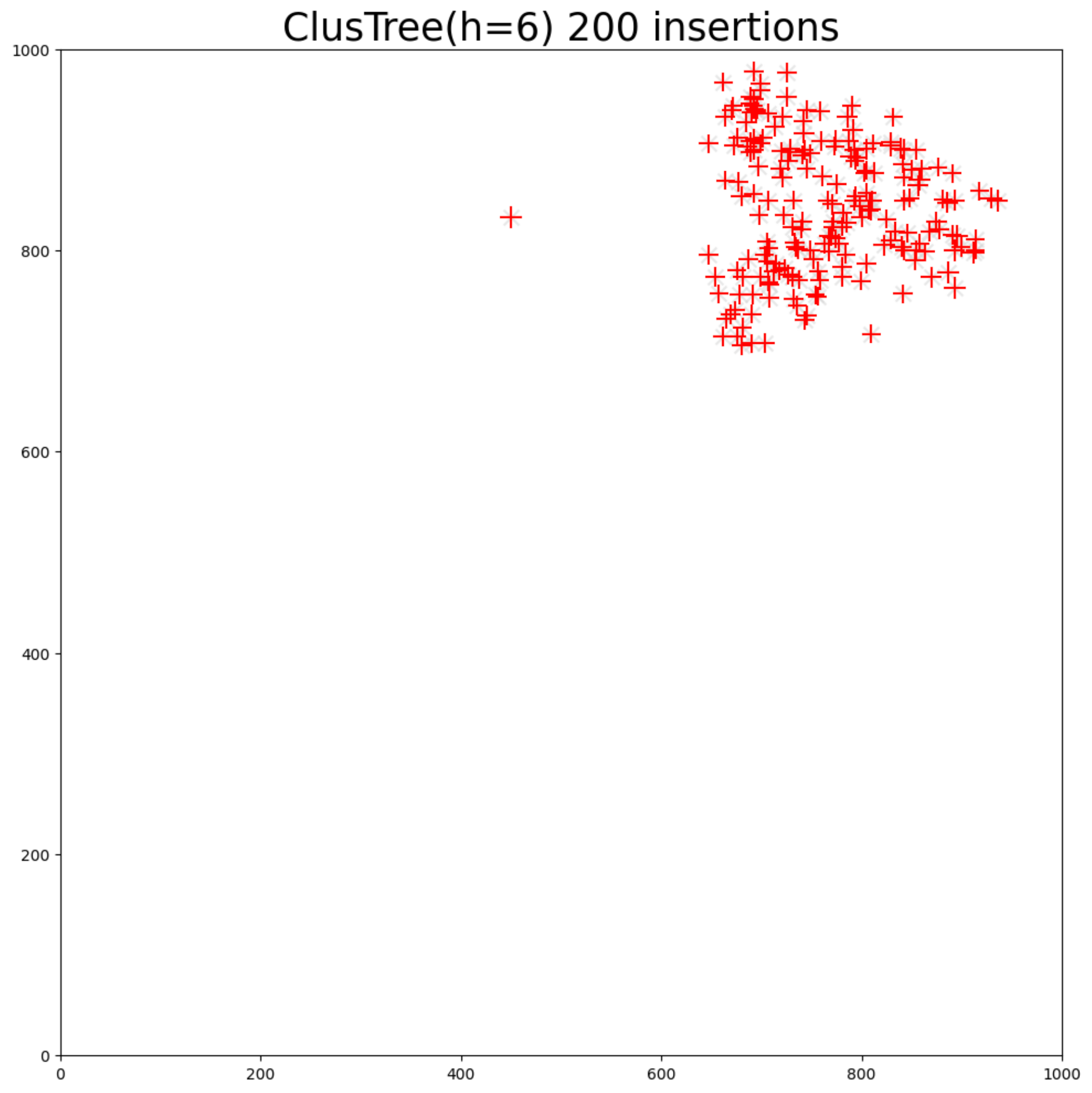}%
            \label{subfig:spreader_clustree_200}%
        }
        \subfloat[]{%
            \includegraphics[width=0.25\linewidth]{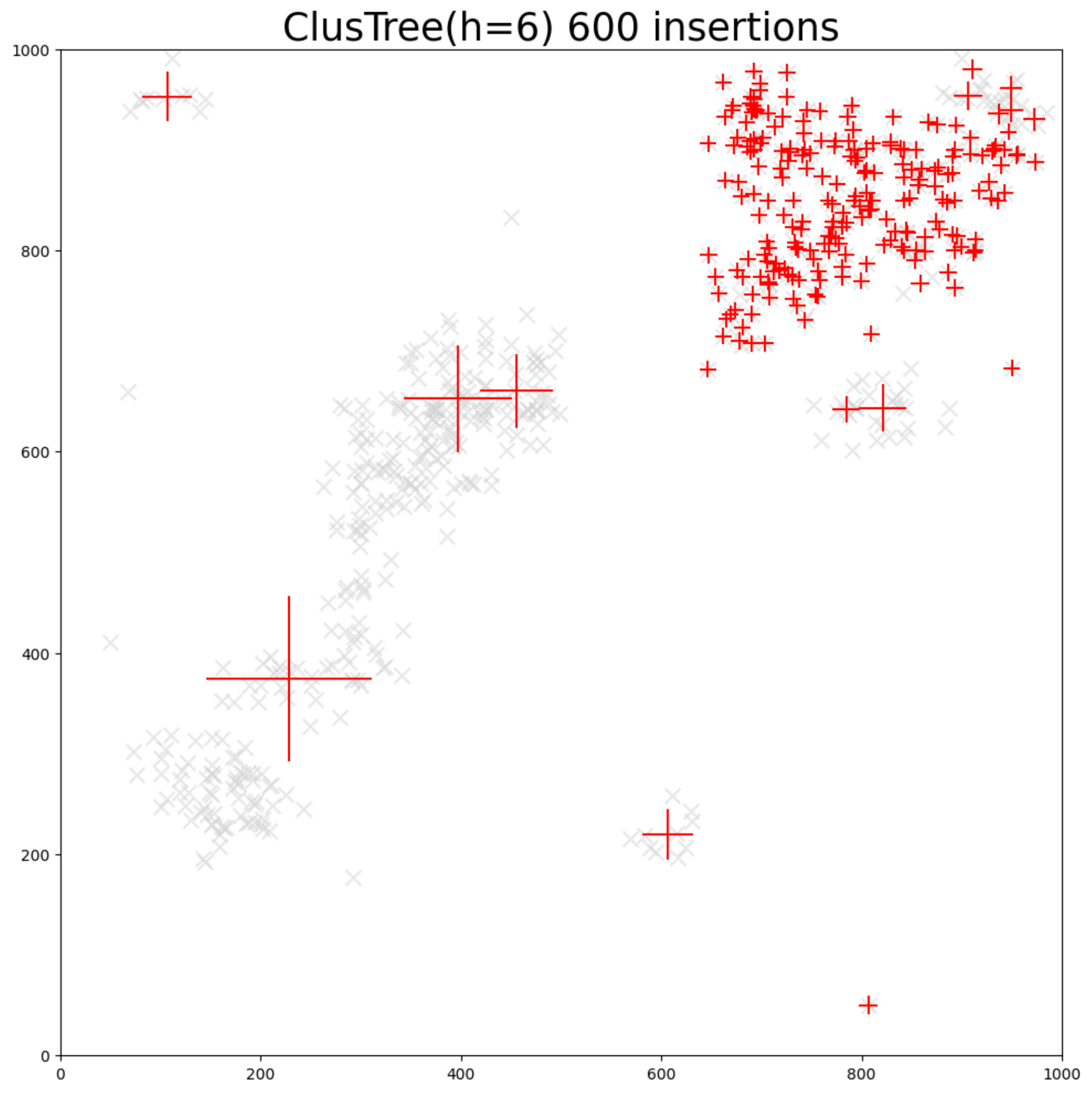}%
            \label{subfig:spreader_clustree_600}%
        }
        \subfloat[]{%
            \includegraphics[width=0.25\linewidth]{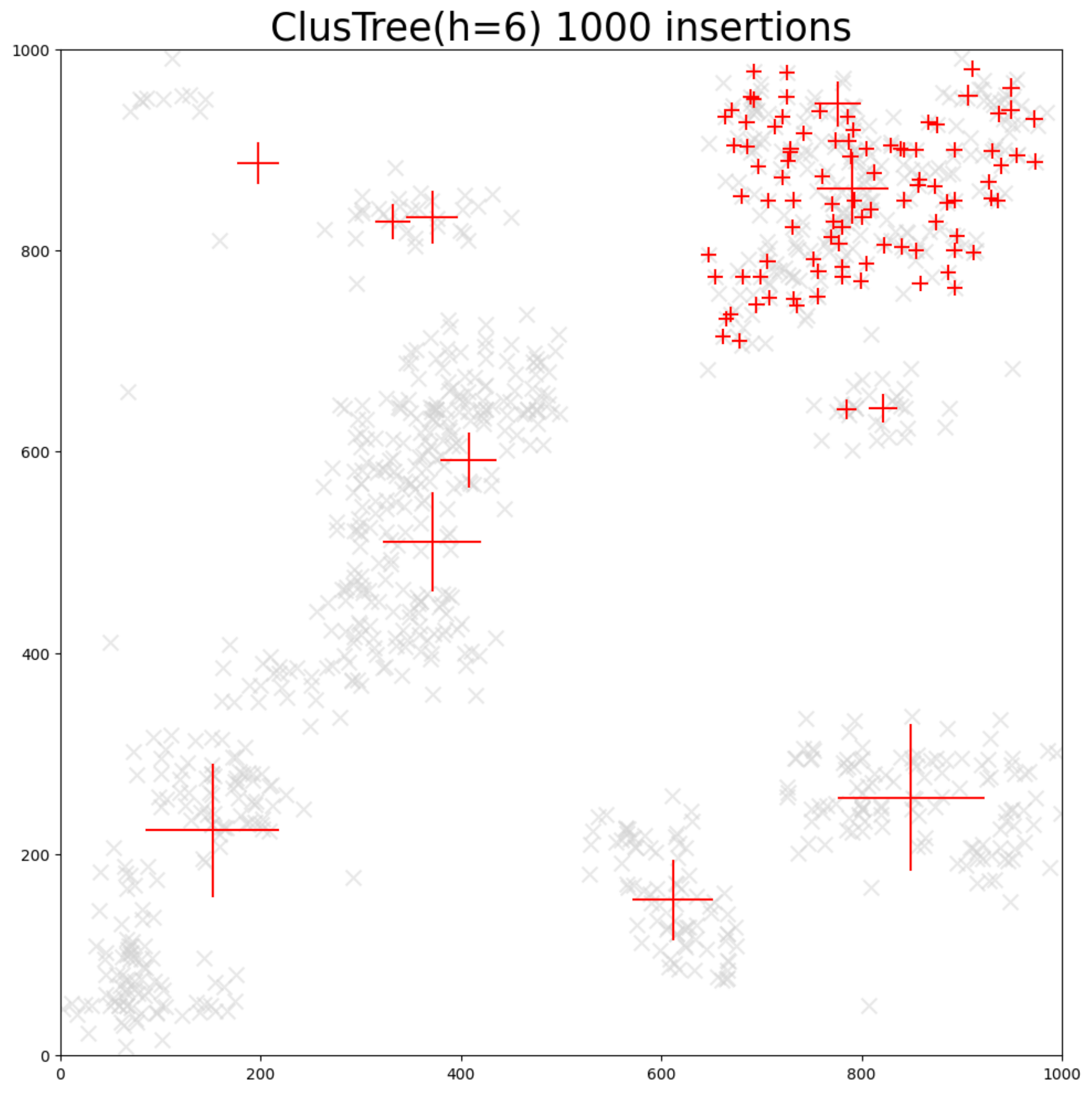}%
            \label{subfig:spreader_clustree_1000}%
        }
        \subfloat[]{%
            \includegraphics[width=0.25\linewidth]{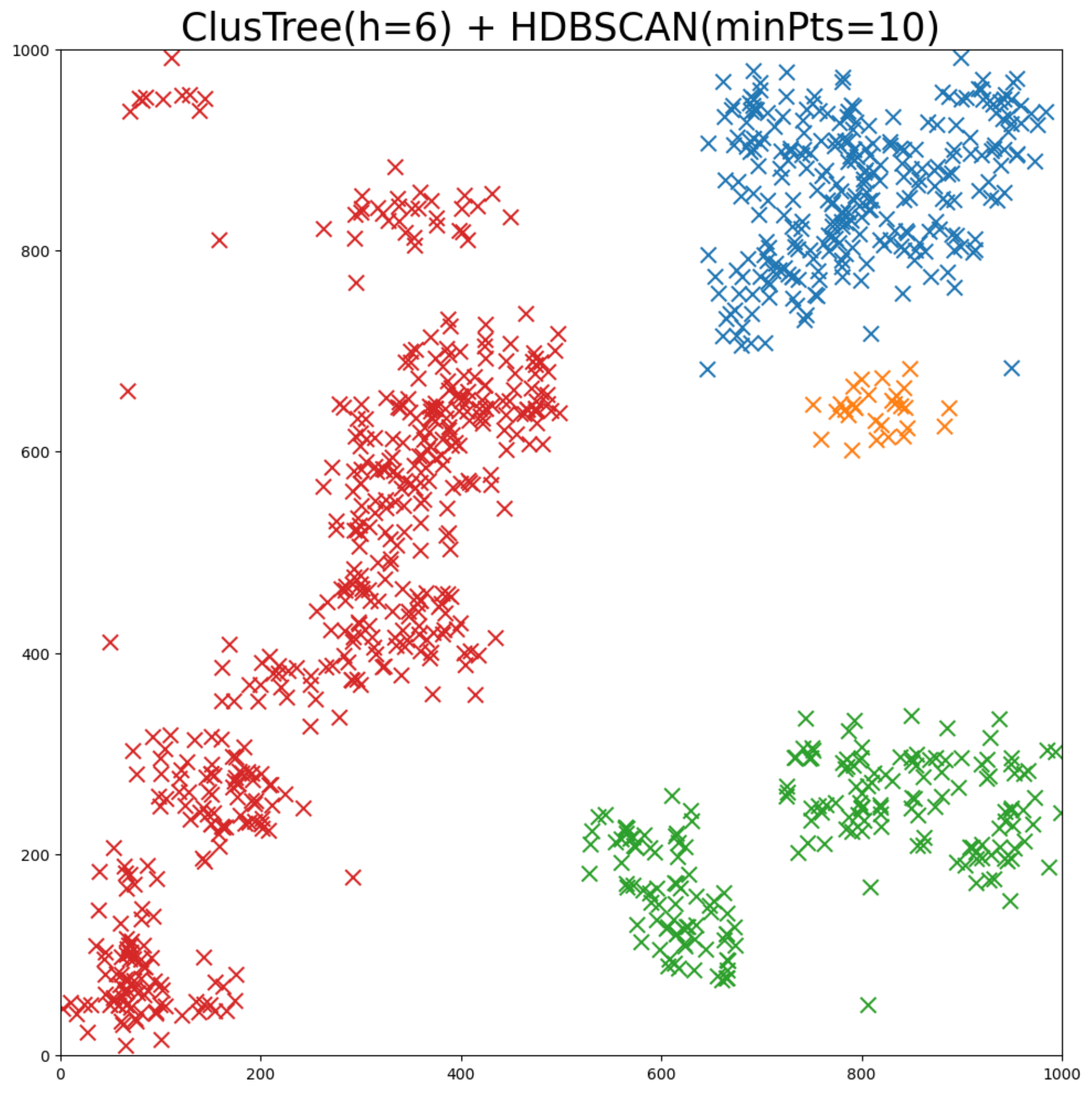}%
            \label{subfig:spreader_clustree}%
        }
        \\
        \subfloat[]{%
            \includegraphics[width=0.25\linewidth]{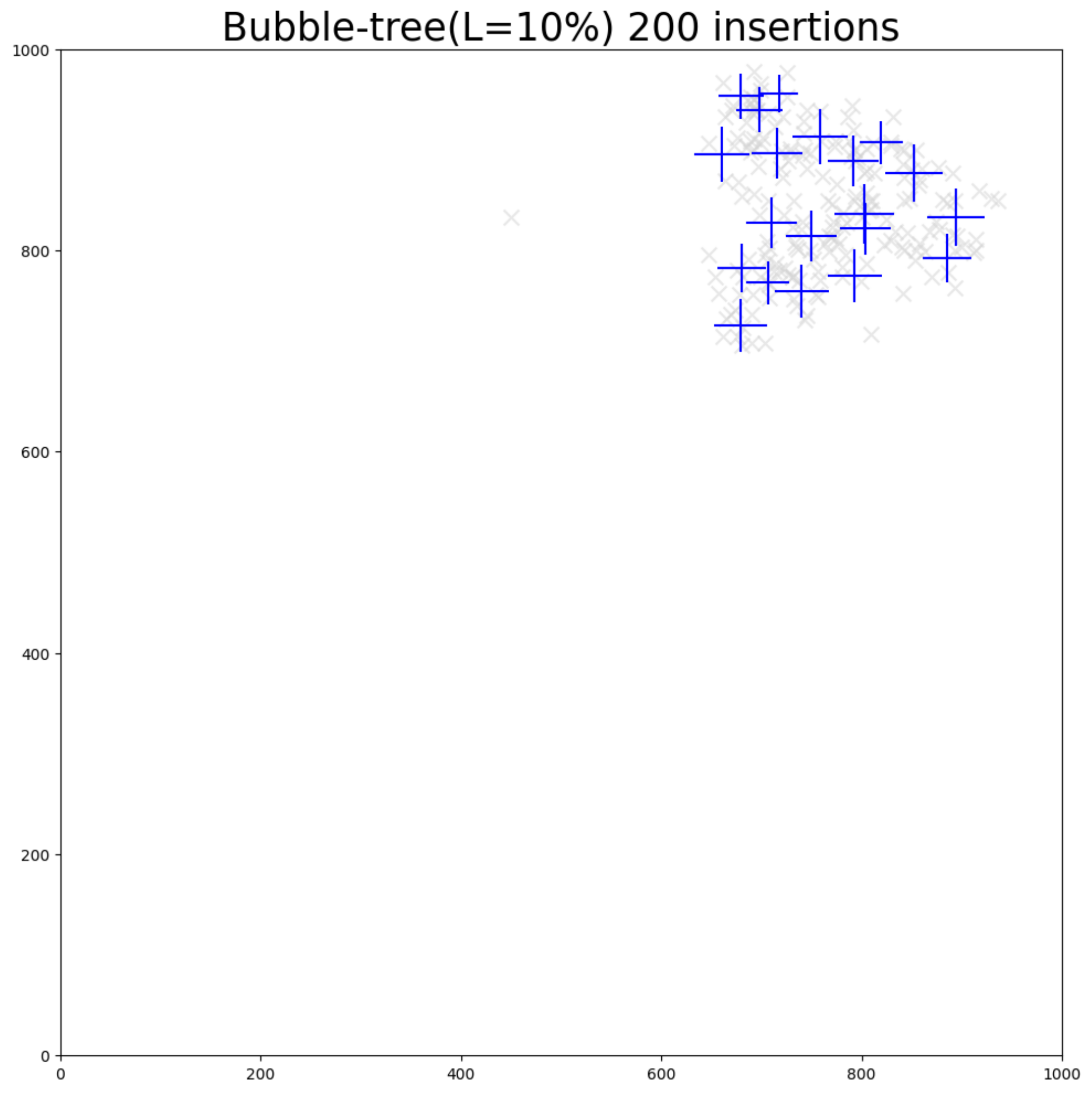}%
            \label{subfig:spreader_bubbletree_200}%
        }
        \subfloat[]{%
            \includegraphics[width=0.25\linewidth]{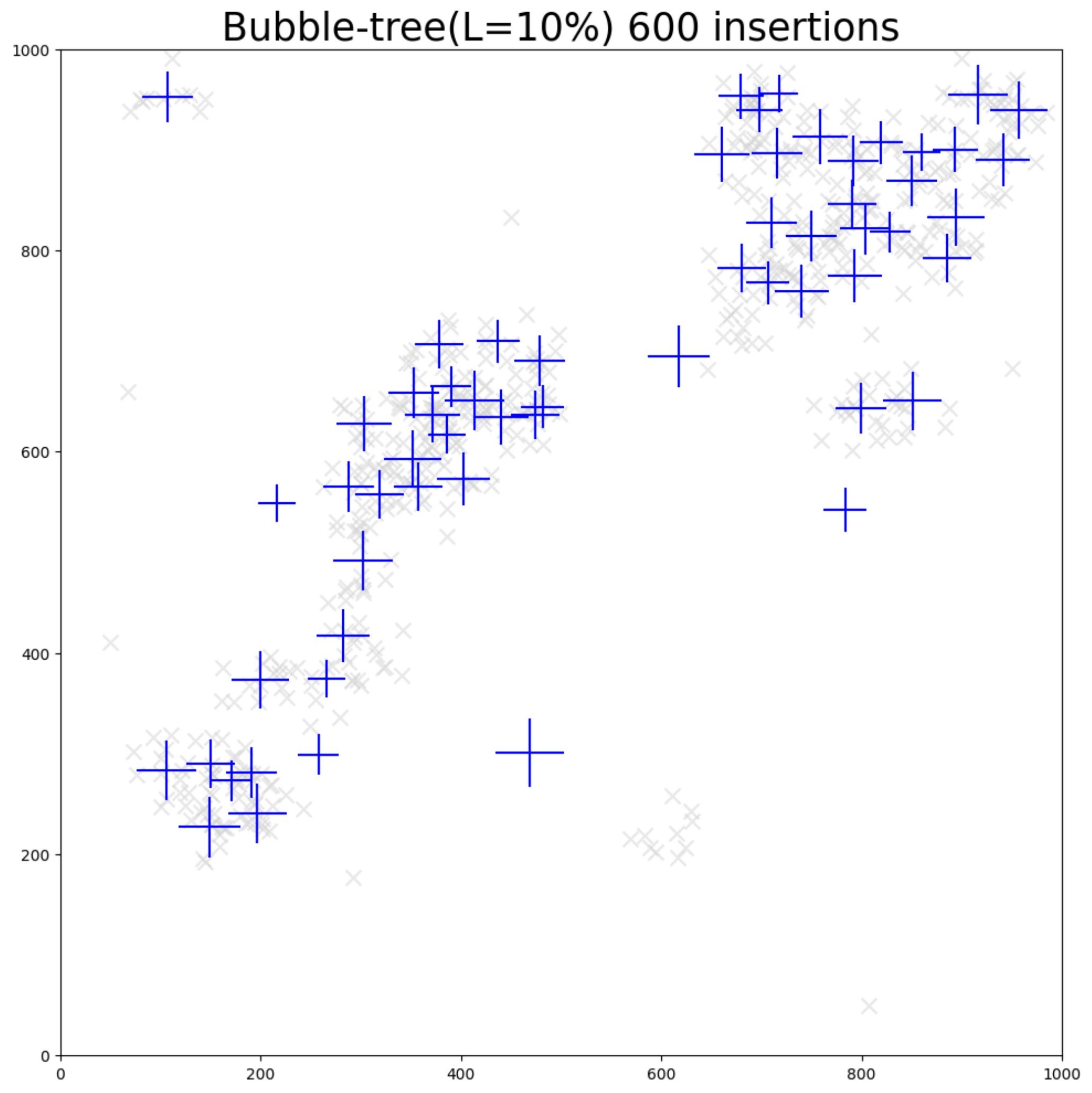}%
            \label{subfig:spreader_bubbletree_600}%
        }
        \subfloat[]{%
            \includegraphics[width=0.25\linewidth]{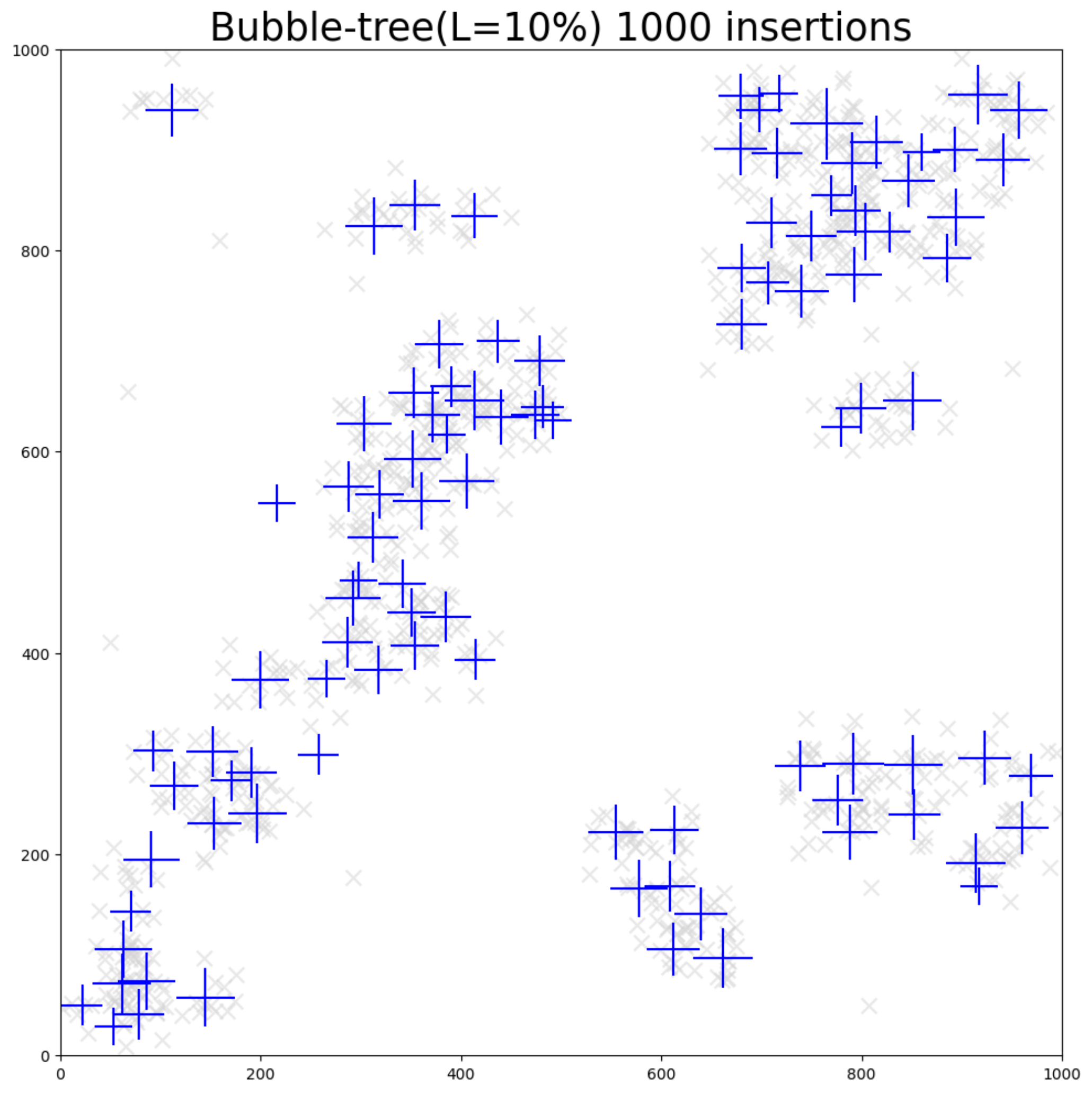}%
            \label{subfig:spreader_bubbletree_1000}%
        }
        \subfloat[]{%
            \includegraphics[width=0.25\linewidth]{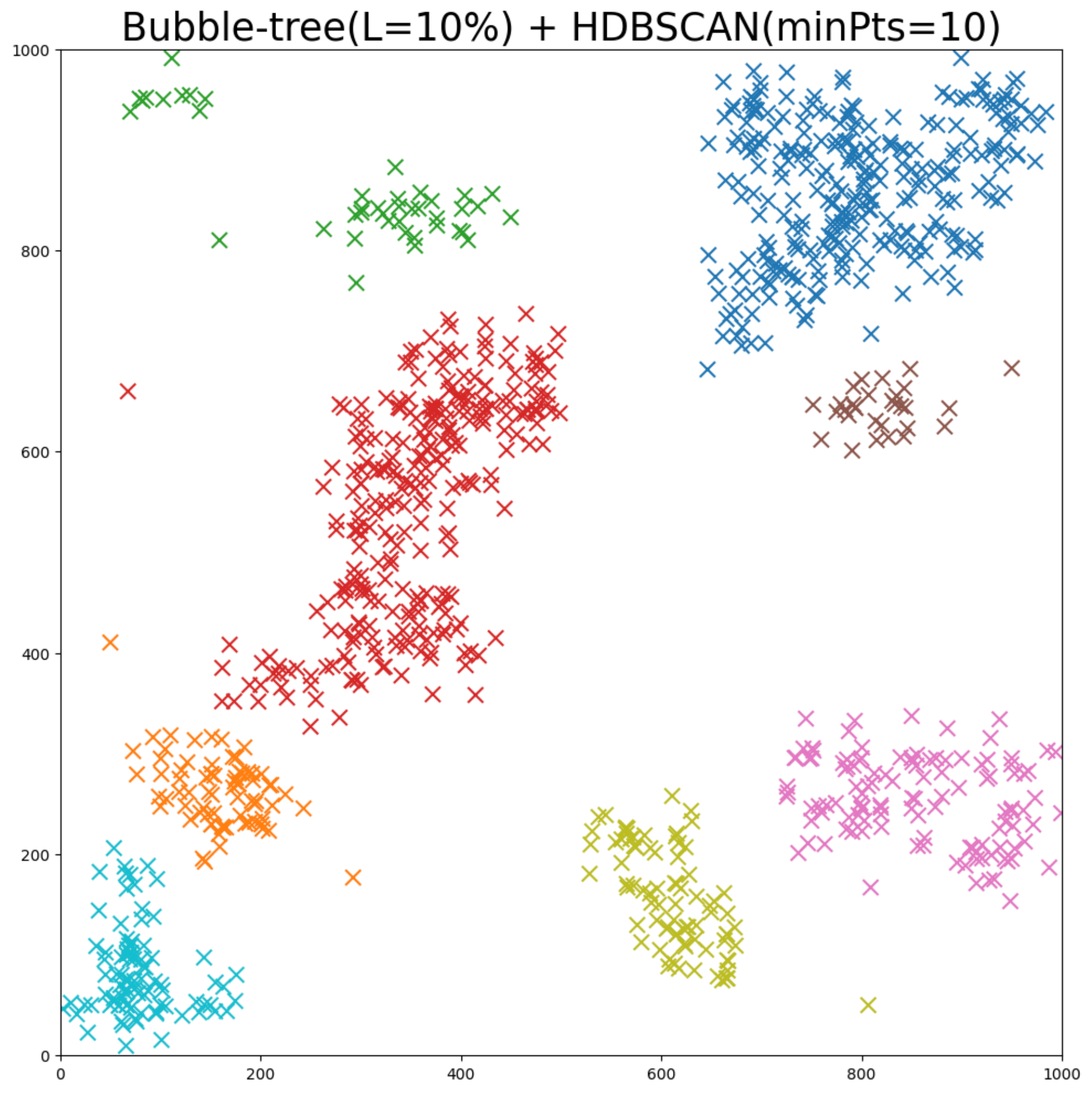}%
            \label{subfig:spreader_bubbletree}%
        }
        \caption{
        Data summarization performed on the 2D example dataset illustrates the differences between ClusTree (a--c) and Bubble-tree (e--g) in incremental settings: original data points (shown in light gray) are inserted incrementally.
        The leaf nodes of the tree structures are shown with plus signs: ClusTree (\textcolor{red}{+}) and Bubble-tree (\textcolor{blue}{+}), the larger the plus sign is, the more data points the leaf node absorbs.
        For the first 200 insertions (a and e), both tree structures compressed the data well, however, next insertions cause ClusTree to create overfilled leaf nodes that has no chance of being split other than inserting more points.
        The resulting leaf nodes from both tree structures were used as compressed data to compute the HDBSCAN clustering results shown in (d) and (h), for ClusTree and Bubble-tree respectively, illustrating how well the tree structures summarize the dynamic data to be used with the clustering algorithm (best viewed in color).
        }
        \label{fig:spreader}
\end{figure*}

\subsection{Dynamic data summarization}
We start by illustrating the differences between fully dynamic data summarization and streaming data summarization, and their impact on summarizing dynamic data to be used in combination with the static algorithm of HDBSCAN.
Here, we compare two baseline data summarization tree structures, Bubble-tree as fully dynamic data summarization and ClusTree for streaming data summarization using the 2D toy example dataset of Seeds.
We configured the maximum height of ClusTree to be 6 and the compression factor of Bubble-tree to be 10\% so that they both approximately have the same number of leaf nodes.
We inserted 100 points iteratively over 10 rounds and plotted the evolving leaf level nodes (micro-clusters) of the tree structures.
After completing all 10 iterations, the micro-clusters were obtained by the leaf level of both tree structures and we applied HDBSCAN on them.

Figures \ref{fig:spreader} (a-c) demonstrates that ClusTree is very susceptible to the data insertion order.
For the first 200 point insertions, ClusTree seems to be summarizing the dynamic data very well as shown in Figure \ref{subfig:spreader_clustree_200}.
However, the next 400 insertions (Figure \ref{subfig:spreader_clustree_600}) forces ClusTree to create bulky leaf level nodes that over-represents the newly inserted points.
After all insertions are performed on ClusTree (Figure \ref{subfig:spreader_clustree_1000}), it started to discard some of the leaf level nodes from the densest region, still many large micro-clusters remain unfairly summarizing the data.
HDBSCAN clustering results obtained using the micro-clusters of ClusTree (Figure \ref{subfig:spreader_clustree}) do not approximate the arbitrary shaped clusters well due to the over-representation of the data.
On the contrary, Bubble-tree adapts well to the nature of fully-dynamic data: newly-made point insertions are balanced to comply with the compression factor while also allowing for \textit{dynamic reorganization} of the previously inserted points as shown in Figures \ref{fig:spreader} (e-g).
Figure \ref{subfig:spreader_bubbletree} shows that HDBSCAN clusters obtained using the micro-clusters of Bubble-tree captures nearly all of the arbitrarily shaped clusters.
This example demonstrates the following key differences between streaming data summarization and dynamic data summarization.
Firstly, dynamic data summarization is \textit{order-independent} in that any arbitrary point in the current data can be deleted, not necessarily in the data insertion order.
This requires us to dynamically reorganize the previously inserted points as the dynamic data evolves.
Secondly, hierarchical spatial clustering requires \textit{balanced} representation of original data points for providing good quality clustering results \cite{data_bubbles}.
This involves adjusting the sizes of the data summaries so that overfilled micro-clusters do not absorb too many data points.
For stream clustering, one of the main requirements is to quickly cluster the newly arriving points within a given time period, so ClusTree is optimal for this purpose \cite{clustree}.
However, ClusTree does not fit into the aforementioned requirements of fully dynamic data: it is order-dependent and creates micro-clusters using an adaptive micro-cluster radius thresholds which causes the creation of overfilled micro-clusters representing too many data points.

\begin{figure}[t!]
\centering
        \subfloat[Pamap]{%
            \includegraphics[width=0.5\linewidth]{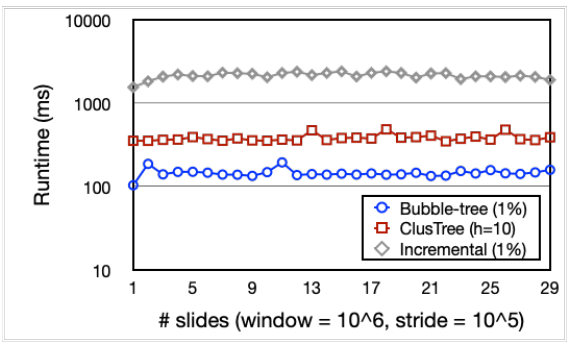}%
            \label{subfig:runtime_pamap}%
        }
        \subfloat[Gauss]{%
            \includegraphics[width=0.5\linewidth]{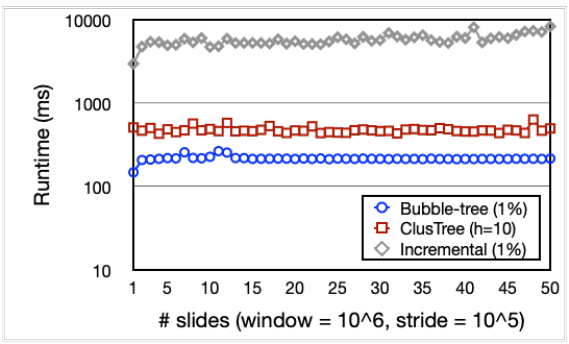}%
            \label{subfig:runtime_gauss}%
        }
        \\
        \subfloat[Chem]{%
            \includegraphics[width=0.5\linewidth]{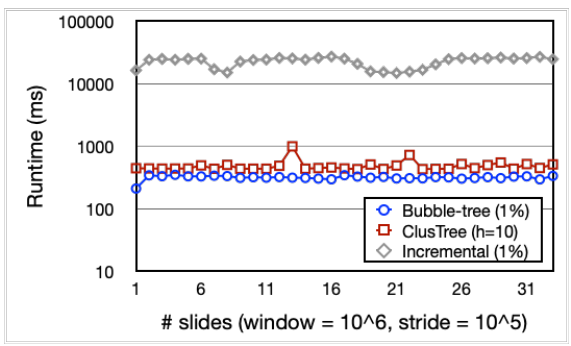}%
            \label{subfig:runtime_chem}%
        }
        \subfloat[Intrusion]{%
            \includegraphics[width=0.5\linewidth]{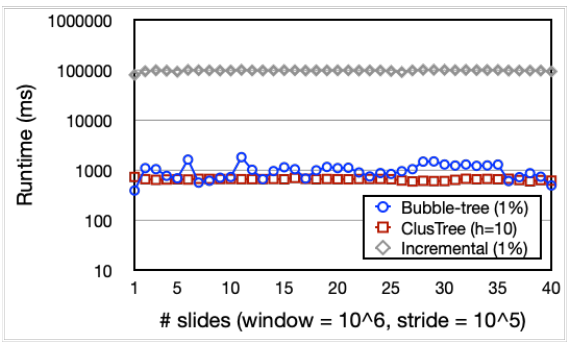}%
            \label{subfig:runtime_intrusion}%
        }
        \caption{
        Running time comparison of data summarization techniques: ClusTree configured with a maximum tree height of 10, roughly equivalent to a 1\% compression rate used in both Bubble-tree and Incremental approaches.
        }
        \label{fig:runtime}
\end{figure}

\begin{figure}
    \centering
    \includegraphics[width=\linewidth]{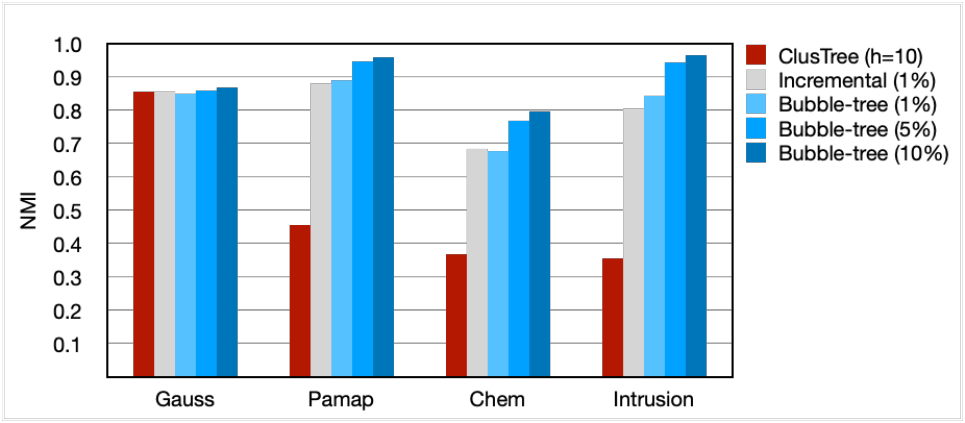}
    \caption{Clustering quality comparison of the data summarization techniques.
     NMI scores reflect how data summarization techniques can achieve similar clustering results with the exact static algorithm of HDBSCAN (best viewed in color).
     }
    \label{fig:quality}
\end{figure}

\subsection{Experiments with real-world data}
Next, we compared the scalability of the baseline algorithms for dynamic data summarization.
The experimental setup involved simulating a sliding window workload to reflect a fully dynamic data environment, where we configured the window size to be $10^6$.
Each iteration of the simulation included a fixed number of data insertions and deletions, where each slide of the window involved the deletion of $10^5$ points of the current data and insertion of $10^5$ new data points to simulate dynamic data flow, ensuring that the baseline algorithms were evaluated under realistic, evolving conditions.
The sliding window experiments were ran following the data generation order of the original datasets.
We evaluated the baseline algorithms in terms of their running time (latency) per slide of the sliding window workload.
We configured Bubble-tree and Incremental approaches with 1\% compression rate, and ClusTree's maximum height was set to 10 to approximately summarize the dynamic data with the same compression rate.
The fanout parameter $M$ of Bubble-tree was set to 10, a common choice for the fanout parameter of dynamic index structures \cite{sstree}.
Figure \ref{fig:runtime} shows an overall report on the running time for the baseline algorithms.

As expected, ClusTree's performance was shown to be optimized for streaming applications, hence, its runtime was balanced for all slides of the window across all experiments.
Incremental approach was the slowest approach as it relies on a straightforward list structure to store its data bubbles.
Overall, the scalability experiments demonstrated that Bubble-tree was very competitive in terms of its runtime per slide of the window, scaling much better that ClusTree and Incremental approaches in the experiments with all datasets except being still competitive in the experiments with Intrusion dataset.

\begin{figure*}[t!]
\centering
        \subfloat[Pamap]{%
            \includegraphics[width=0.25\linewidth]{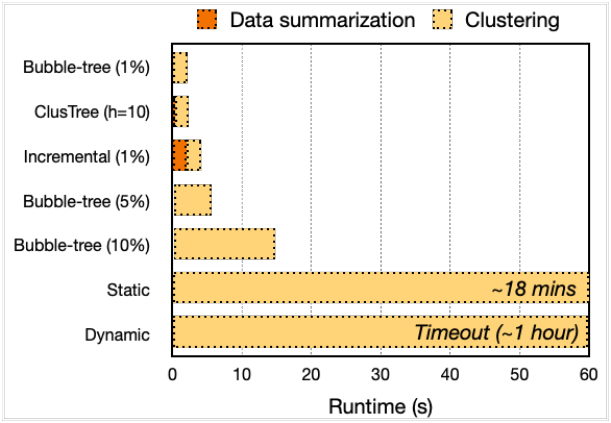}%
            \label{subfig:latency_pamap}%
        }
        \subfloat[Gauss]{%
            \includegraphics[width=0.25\linewidth]{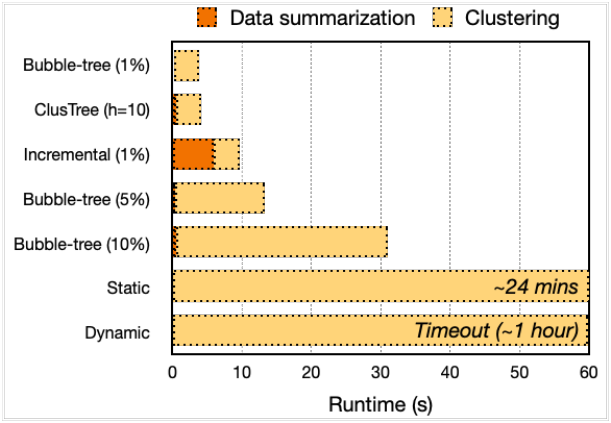}%
            \label{subfig:latency_gauss}%
        }
        \subfloat[Chem]{%
            \includegraphics[width=0.25\linewidth]{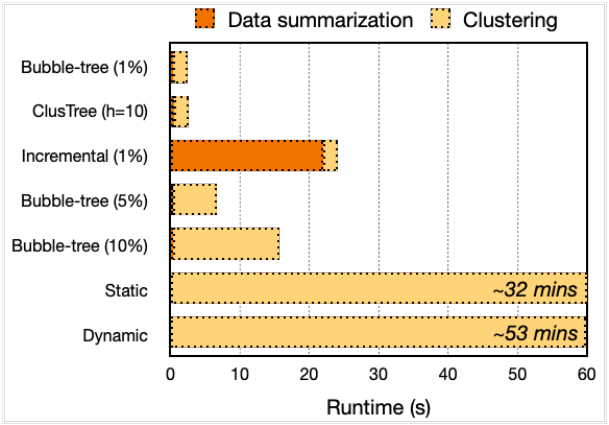}%
            \label{subfig:latency_chem}%
        }
        \subfloat[Intrusion]{%
            \includegraphics[width=0.25\linewidth]{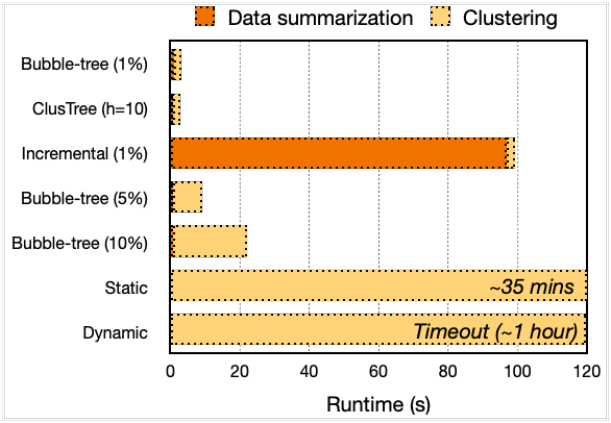}%
            \label{subfig:latency_intrusion}%
        }
        \caption{
        Running time comparison of three data summarization techniques (Bubble-tree, ClusTree, and Incremental) applied to the clustering algorithm, alongside the exact dynamic (Dynamic) and the static HDBSCAN (Static) algorithms, highlighting the scalability benefits of the data summarization techniques (best viewed in color).
        }
        \label{fig:latency}
\end{figure*}

On each slide of the workload, we measured the clustering quality obtained from the offline component of the baseline algorithms using the Normalized Mutual Information (NMI) score from the scikit learn library \cite{scikit}.
NMI is robust for comparing clustering results with noise by measuring the agreement between the predicted flat clustering results obtained by the offline component of the data summarization techniques in comparison to the clustering results obtained by the static algorithm, scaled between 0 (no mutual information) and 1 (perfect correlation).
Figure \ref{fig:quality} includes an average NMI scores achieved by the baseline algorithms when compared to the clustering results obtained by the static algorithm.
Similar to our previous analysis of ClusTree being unsuitable for dynamic data summarization, it showed the lowest quality clustering results.
The only exception was that ClusTree was able to summarize the Gaussian clusters very well, since it is optimized for detecting circularly-shaped k-means type clusters.
However, ClusTree's data summarization performed poorly on real-world datasets that contains arbitrarily shaped clusters, as shown with Pamap, Chem, and Intrusion datasets.
Incremental approach for dynamic data summarization provided high quality clustering results, confirming the results in the literature \cite{maintain_fixed_data_bubbles}, but the approach suffered from scalability issues as the dimensionality of the data increases (Figure \ref{fig:latency}).

Across all experiments, Bubble-tree demonstrated high quality clustering results even at 1\% compression rate.
We additionally include the experimental results with Bubble-tree configured with 5\% and 10\% compression rates.
Figure \ref{fig:latency} shows an overall report on the scalability benefits achieved by the baseline algorithms.
The results demonstrated that Bubble-tree can support higher compression rates if desired by the user for higher quality clustering results.
Compared to the exact dynamic algorithm and the static algorithms of HDBSCAN, for example, we can summarize the dynamic data with 10\% compression rate and achieve good quality clustering results under a minute runtime threshold.
This is especially evident from the experiments with Intrusion dataset: the exact dynamic could not finish in one hour, and the static algorithm took 35 minutes, whereas Bubble-tree configured with 10\% compression rate completed each slide of the window spending on average 20 seconds.

\section{Related work}
Recent advances in hierarchical density-based clustering have focused on the sequential and parallel computation of the non-incremental MST for HDBSCAN.
HDBSCAN* \cite{accelerated_hdbscan} is an accelerated HDBSCAN approach that proposes a sequential MST algorithm based on a dual-tree framework called Dual-tree Boruvka \cite{dualtreeboruvka}.
A parallel algorithm to compute the Euclidean MST \cite{fast_parellel_hdbscan}, based on well-separated pair decomposition \cite{wspd}, has been adapted for HDBSCAN.
Several research directions include the use of space partitioning tree structures \cite{accelerated_hdbscan} or kNN graph approximation \cite{approx_hdbscan_knn_graph} and hierarchical navigable small world \cite{fishdbc} to approximate the MST of HDBSCAN.
Relative neighborhood graphs \cite{rng} were utilized for HDBSCAN \cite{rng_for_hdbscan} to explore multiple values of the density parameter of HDBSCAN.

HASTREAM\cite{ihastream} is a streaming clustering algorithm that adapts the concepts of mutual reachability and core distances based on a micro-cluster structure in a damped window model, assigning more importance to recent data in a streaming environment.
FISHDBC \cite{fishdbc} proposed an approximate incremental-only MST algorithm for HDBSCAN with which MST computation can be delayed w.r.t. the corresponding update parameter and the MST can be  recomputed using the edges of the current MST in addition to candidate edges collected from new insertions made to an approximate index structure.
However, when the update parameter is set to 1, FISHDBC computes the MST from scratch for every point insertion, meaning that the time complexity is nearly identical to that of the non-incremental Dual-tree Boruvka approach \cite{dualtreeboruvka}.
Regardless of the characteristics of these algorithms, whether they are exact or approximate and/or sequential or parallel, a fully dynamic algorithm for HDBSCAN remains a major challenge, especially for applications that work with dynamic data without bias regarding the order of the data points allowing arbitrary point insertions and deletions.

\section{Conclusion}
\label{sec:conclusion}
In this paper, we studied the problem of dynamic hierarchical spatial clustering.
Firstly, we developed an exact algorithm using dynamic index and tree structures which performed well on small changes in the dynamic data.
Analyzing the hardness of such exact algorithm for modern dynamic data applications involving unpredictable changes in the dynamic data, we considered data summarization techniques to mitigate the scalability issues of both the static and exact dynamic algorithms.
We discussed the applicability of existing data summarization techniques for fully dynamic settings.
To achieve both scalable and high quality data summarization in the fully dynamic settings, we presented a dynamic tree structure, called \textit{Bubble-tree}, for maintaining a compressed form of fully dynamic data for a given compression factor.
Bubble-tree is capable of maintaining high quality data summaries offering rapid update operations in the fully dynamic settings.
Experimental analysis shows that our proposals can maintain high quality data summarization of dynamic data, closely competitive for the existing techniques while achieving orders of magnitude speedups in running time.
As future research, we will adapt our proposals for popular hierarchical spatial clustering frameworks in the Rust ecosystem by open-sourcing the code and develop anytime capability for handling unpredictable fully dynamic data workloads.

\bibliographystyle{abbrv}
\bibliography{references}

\appendix
\section{Reverse Nearest Neighbors Search}
\label{appendix:rknn}
Reverse k Nearest Neighbors (RkNN) query searches for data points that will have the query point as one of their k nearest neighbors \cite{rknn}.
The main idea of a RkNN query is related to the use of the cached core distances of data points to obtain more pruning power during node traversals.
Algorithm \ref{alg:rknn} in the Appendix \ref{appendix:rknn} includes an RkNN query presented in earlier work \cite{rdnn}.
Given query point $p$, the algorithms starts with the root of the spatial index.
A traversal of tree node $A$ can be pruned (Lines 1-2) if the maximum core distance value $cd_{max}(A)$ of the descendant points of node $A$ do not reach $p$, meaning that none of the descendant points registers $p$ as one of their $k$ nearest neighbors.
When a tree traversal reaches a leaf node, it checks for every point in the leaf node to find the RkNNs of $p$ (Lines 3-6).
Otherwise, the algorithm recursively traverses down in the tree hierarchy by visiting each child of the current node (Lines 7-8).
For the RkNN query to be efficient, $cd_{max}(A)$ are kept up to date when the dynamic spatial index undergoes point insertions and deletions.

\SetCommentSty{mycommfont}
\begin{algorithm}
\KwIn{$p$: point, $A$: node}
\KwOut{$R_{minPts}$: reverse neighborhood of $p$}
    \SetKwFunction{func}{ReverseKnn}
    \func{$p$, $A$} \\
        \nl \If{$d_{min}(p, A) \geq cd_{max}(A)$}{
        \nl     \Return
            }
        \nl \uIf{$A$ is a leaf node}{
                \nl \For{$q \in A.points$}{
                    \nl \If{$cd(q) > d(p, q)$}{
                    \nl     $R_{minPts} \gets R_{minPts} \cup q$
                        }
                }
            }
            \uElse {
                \nl \For{$child \in A.children$}{
                    \nl \func{$p$, $child$}
                }
            }
    \caption{RkNN query on point $p$}
\label{alg:rknn}
\end{algorithm}

\SetCommentSty{mycommfont}
\begin{algorithm}[t!]
    \SetKwFunction{func}{DualTreeBoruvka}
    \SetKwFunction{updatetree}{UpdateTree}
    \SetKwFunction{unionfind}{UnionFind}
    \SetKwFunction{findnext}{FindComponentNeighbors}
    \SetKwFunction{queryrange}{QueryRange}
    \KwIn{$A$: spatial index, $T$: MSF}
    \KwOut{$T$: MST}
    \func{$A$, $T$} \\
        \nl $U \gets \unionfind(T)$ \\
        \nl \updatetree{$A.root$, $U$} \\
        \nl \While{$|T| < N - 1$}{
            \nl $E \gets $ \findnext{$A.root$, $A.root$} \\
            \nl \For{$(p, q) \in E$}{
                \tcp{Merge components}
                \nl \If{$U.find(p) \neq U.find(q)$}{
                    \nl $U.union(p, q)$ \\
                    \nl $T \gets T \cup (p, q)$ \\
                    }
                }
            \nl \updatetree{$A.root$, $U$} \\
        }
    \caption{Compute Boruvka MST}
\label{alg:dual_tree_boruvka}
\end{algorithm}

\SetCommentSty{mycommfont}
\begin{algorithm}[t!]
\KwIn{$Q$: node, $R$: node}
\KwOut{$E$: a set of edges}
    \SetKwFunction{func}{FindComponentNeighbors}
    \func{$Q$, $R$} \\
        \nl \If{$Q$ and $R$ are in the same component}{
        \nl     \Return
            }
        \nl \If{$d_{min}(Q, R) \geq d(Q)$}{
        \nl     \Return \tcp{Prune if there is a better neighbor}
            }
        \nl \uIf{$Q$ and $R$ are both leaf nodes}{
                \nl \For{$q \in Q.points$}{
                    \nl \For{$r \in R.points$}{
                       \nl \If{$d(C_q) > d(q, r)$}{
                                \nl $d(C_q) = d(q, r)$ \\
                                \nl $e(C_q) = (q, r)$ \\
                                \nl $E \gets E \cup (q, r)$
                            }
                    }
                }
                \nl $d(Q) = \max_{q \in Q} \{ d(C_q) \}$
            }
        \nl \uElseIf{$Q$ is a leaf node} {
                \nl \For{$child \in R.children$}{
                    \nl \func{$Q$, $child$}
                }
            }
        \nl \uElseIf{$R$ is a leaf node} {
                \nl \For{$child \in Q.children$}{
                    \nl \func{$child$, $R$}
                }
            }
        \nl \uElse{
                \nl \For{$q \in Q.children$}{
                    \nl \For{$r \in R.children$}{
                        \nl \func{$q$, $r$}
                    }
                }
        }
    \caption{Find next component edges}
\label{alg:find_next_neighbors}
\end{algorithm}

\section{Dual-tree Boruvka}
\label{appendix:dual_tree_boruvka}
Dual-tree Boruvka \cite{dualtreeboruvka} makes use of a spatial index to compute the MST of the data points efficiently.
Algorithm \ref{alg:dual_tree_boruvka} presents the pseudocode obtained from the literature \cite{dualtreeboruvka}, modified here to support any spatial index, as presented in \cite{faster_boruvka}.
We also modified the original algorithm to have a minimum spanning forest $T$ as input; for example, if $T$ has no edges, there are $N$ components, meaning that the algorithm computes the MST from scratch.
Initially, the algorithm builds a union find $U$ of the current connected components for the given minimum spanning forest $T$ (Line 1).
Next, the spatial index is updated with $U$ (Line 2), that is, the $UpdateTree$ subroutine traverses the spatial index to update connected components.
Tree node $A$ represents a single component if all of its children are in the same component.
The algorithm repeatedly runs $FindComponentNeighbors$ (Algorithm \ref{alg:find_next_neighbors}) to find shortest outgoing edges of the current components until only one component remains (Lines 4-8).

Algorithm \ref{alg:find_next_neighbors} shows the pseudocode used to find the next outgoing edges of the components.
The subroutine is called with the spatial index's root as both query $Q$ and reference $R$ nodes (see Algorithm \ref{alg:dual_tree_boruvka} Line 4).
There are two possible pruning cases: if both $Q$ and $R$ are in the same component (Lines 1-2) or the current shortest outgoing edge of $Q$ is shorter than the minimum distance between $Q$ and $R$ (Lines 3-4).
Otherwise, every combination of child pairs is recursively traversed (Lines 12-21) unless the base case involves both nodes being leaf nodes (Lines 5-11).
In the base case, where both $Q$ and $R$ are leaf nodes, every combination of the corresponding child points $(q, r)$ is checked to determine if the pair can be a lighter edge than the current outgoing candidate edge (Lines 8-10).

\section{Dynamic hierarchical spatial clustering}
\label{appendix:exact_dynamic}
To realize an exact dynamic hierarchical spatial clustering, we assume an online maintenance of MST of HDBSCAN where the offline phase extracts the clustering hierarchy (dendrogram) at a user request.
We use the following data structures: a dynamic spatial index, SS-tree \cite{sstree} for maintaining core distance information and a link-cut tree \cite{link_cut_tree} for maintaining the MST.
Both insertion and deletion algorithms synchronizes the core distance information using kNN \cite{knn} and RkNN queries \cite{rknn} using a modified version of standard dynamic index operations \cite{rdnn} (Appendix \ref{appendix:rknn}).

The insertion of point $p$ starts by performing kNN query \cite{knn} and RkNN query (Algorithm \ref{alg:rknn}) with query point $p$ on the dynamic index $A$ to find $N_{minPts}(p)$ and $R_{minPts}(p)$ respectively.
While $N_{minPts}(p)$ is used to compute $p$'s core distance $cd(p)$, each of $p$'s reverse neighbors in $R_{minPts}(p)$ recalculates their core distances by including $p$ as a neighbor.
Following this, a modified insertion algorithm of dynamic index is called to insert point $p$ into $A$ which synchronizes the updated neighborhood information of the dynamic index \cite{rdnn}.
Next, the newly-inserted edges of the underlying mutual reachability graph, $E_{inserted}$, are created by connecting $p$ to the existing points. 
The edges with modified weights are found by connecting each of $p$'s reverse neighbors in $R_{minPts}(p)$ to their nearest neighbors.
Finally, the insertion of the edges involves updating the MST using a link-cut tree operations \cite{cattaneo2010maintaining} by checking the edges of $E_{inserted}$ and $E_{modified}$ (Eq. \ref{eq:mst_insertion}): an edge is inserted into the MST if it does not create a cycle or can replace the maximum weighted edge on the cycle.
The pseudocode for inserting edge $(u, v)$ is illustrated in Algorithm \ref{alg:insertion} (Lines 6-12).
First, we check if points $u$ and $v$ are already connected in $T$. If they are connected, the insertion of $(u, v)$ creates a cycle; otherwise, we can safely insert the edge into $T$.
With regard to cycle formation, we find the heaviest edge $(x, y)$ in the cycle to check if $(u, v)$ can replace $(x, y)$ by comparing their weights.
If this is the case, we then cut $(x, y)$ and insert $(u, v)$ into $T$.
The resulting MST $T$ can be used at a user request for extracting the clustering hierarchy or flat clusters results.

The deletion algorithm starts by calling a standard deletion algorithm on the dynamic spatial index $A$ to delete point $p$.
RkNN query (Algorithm \ref{alg:rknn}) is performed on the dynamic spatial index $A$ to find the RkNNs of $p$, $R_{minPts}(p)$.
The reverse neighbors in $R_{minPts}(p)$ recalculate their core distances by calling kNN queries on the dynamic index which can be done in batch efficiently \cite{rdnn}.
Next, the MST edges incident to point $p$ and any of its reverse neighbors in $R_{minPts}(p)$ are deleted from the MST $T$, resulting in a minimum spanning forest.
Finally, Dual-tree Boruvka \cite{dualtreeboruvka} algorithm (Algorithm \ref{alg:dual_tree_boruvka}) is called to reconnect the disconnected components of the forest which returns the updated MST $T$ for further usage by the user.

\SetCommentSty{mycommfont}
\begin{algorithm}[t!]
\caption{Insert point $p$}
\label{alg:insertion}
\SetKwFunction{knn}{Knn}
\SetKwFunction{rknn}{ReverseKnn}
\KwIn{$T$: current MST, $A$: spatial index, $p$: point}
\KwOut{$T$: updated MST}
    \SetKwFunction{func}{Insert}
    \func{$T$, $A$, $p$} \\
    \tcp{Update core distance information}
    \nl $N_{minPts}(p) \gets \knn(p, A)$ \\
    \nl $cd(p) \gets \max_{p' \in N_{minPts}(p)} d(p, p')$ \\
    \nl $R_{minPts}(p) \gets \rknn(p, A)$ \\
    \nl \For{$r \in R_{minPts}(p)$}{
        \nl $cd(r) \gets \max_{N_{minPts}(r)} d(r, r')$ \\
    }
    \nl Insert $p$ into the spatial index $A$ \\
    \tcp{Create candidate edges}
    \nl $E_{inserted} \gets \{ (p, p') \mid p' \in A.points \}$ \\
    \nl $E_{modified} \gets \{ (r, r') \mid r \in R_{minPts}(p) \mid r' \in N_{minPts}(r)\}$ \\
    \tcp{Update MST $T$:}
    \nl \For{$(u, v) \in E_{inserted} \cup E_{modified}$}{
        \nl         \If{$\Connected(u, v)$}{
        \nl             $(x, y) \gets \FindMax(u, v)$ \\
        \nl             \If{$d_m(x, y) \leq d_m(u, v)$}{
        \nl                 \Continue 
                        }
        \nl             $\Cut(x, y)$ \tcp{Prevent cycle formations}
                    }
        \nl     $\Link(u, v)$ \\
    }
\end{algorithm}

\SetCommentSty{mycommfont}
\begin{algorithm}[t!]
\caption{Delete point $p$}
\label{alg:deletion}
\SetKwFunction{knn}{Knn}
\SetKwFunction{rknn}{ReverseKnn}
\SetKwFunction{boruvka}{DualTreeBoruvka}
\KwIn{$T$: current MST, $A$: spatial index, $p$: point}
\KwOut{$T$: updated MST}
    \SetKwFunction{func}{Delete}
    \func{$T$, $A$, $p$} \\
    \tcp{Update core distance information}
    \nl Delete $p$ from the spatial index $A$ \\
    \nl $R_{minPts}(p) \gets \rknn(p, A)$ \\
    \nl \For{$r \in R_{minPts}(p)$}{
        \nl $cd(r) \gets \max_{N_{minPts}(r)} d(r, r')$ \\
    }
    \tcp{Delete MST edges}
    \nl $E_{deleted} \gets \{ (p, p') \in T \}$ \\
    \nl $E_{modified} \gets \{ (r, r') \in T \mid r \in R_{minPts}(p) \}$ \\
    \nl $T \gets T \setminus (E_{deleted} \cup E_{modified})$ \\
    \tcp{Update MST $T$:}
    \nl $\boruvka(A, T)$
\end{algorithm}

\end{document}